\documentclass[manuscript]{aastex}
\usepackage{natbib}
\usepackage{color}
\usepackage{graphicx}
\usepackage{amsmath}
\usepackage{morefloats}
\usepackage{multirow}
\newcommand\scalemath[2]{\scalebox{#1}{\mbox{\ensuremath{\displaystyle #2}}}}
\begin{document}
\title{Maximum Likelihood Foreground Cleaning for Cosmic Microwave Background Polarimeters in the Presence 
  of Systematic Effects}

\author{C.Bao\altaffilmark{1}, C.Baccigalupi\altaffilmark{2}, B.Gold\altaffilmark{3}, 
        S.Hanany\altaffilmark{1}, A.Jaffe\altaffilmark{4}, R. Stompor\altaffilmark{5}}
\altaffiltext{1}{University of Minnesota School of Physics and Astronomy,
Minneapolis, MN 55455}
\altaffiltext{2}{SISSA, Astrophysics Sector, via Bonomea 265, Trieste 34136, Italy}
\altaffiltext{3}{Hamline University, Saint Paul, MN 55104}
\altaffiltext{4}{Imperial College, London, SW72AZ, England, United Kingdom}
\altaffiltext{5}{Laboratoire Astroparticule et Cosmologie (APC), 75205 Paris Cedex 13, France}

\begin{abstract}

We extend a general maximum likelihood foreground estimation for cosmic microwave background polarization 
data to include estimation of instrumental systematic effects. We focus on two particular effects: frequency 
band measurement uncertainty, and instrumentally induced frequency dependent polarization rotation. We 
assess the bias induced on the estimation of the $B$-mode polarization signal by these two systematic 
effects in the presence of instrumental noise and uncertainties in the polarization and spectral index of 
Galactic dust. Degeneracies between uncertainties in the band and polarization angle calibration 
measurements and in the dust spectral index and polarization increase the uncertainty in the extracted CMB 
$B$-mode power, and may give rise to a biased estimate. We provide a quantitative assessment of the 
potential bias and increased uncertainty in an example experimental configuration. For example, 
we find that with 10\% polarized dust, tensor to scalar ratio of $r=0.05$, and the instrumental 
configuration of the EBEX balloon payload, the estimated CMB $B$-mode power spectrum 
is recovered without bias when the frequency band measurement has 5\% uncertainty or less,  
and the polarization angle calibration has an uncertainty of up to 4$^{\circ}$. 
\end{abstract}

\keywords{cosmic microwave background --- instrumentation: polarimeters ---
          methods: data analysis}

\section{Introduction}
The inflationary paradigm posits that the universe underwent a period of exponential expansion within its 
first fraction of a second. One of the generic predictions of inflation is the production of stochastic
gravitational wave background and the imprint of a $B$-mode pattern in the polarization of the 
cosmic microwave background~(CMB) radiation on large angular scales~\citep{kamionkowski97a,seljak97}. The 
level of the inflationary gravitational wave $B$-mode signal encodes information about the inflationary 
energy scale and is characterized by a tensor-to-scalar ratio~$r$ which quantifies the relative strength of 
gravity waves and density perturbations produced during inflation. The current upper limit from $B$-mode 
observation alone is $r<0.12$~\citep{bkp2015, planck2015parameters}. On small angular scales there 
is more dominant source of $B$-mode polarization: lensing of CMB photons by the large scale structure of the 
universe, which converts CMB $E$-mode to $B$-mode~\citep{zaldarriaga98}.

According to recent observations, polarized Galactic thermal dust emission is a significant source of 
contamination for the inflationary $B$-mode signal~\citep{Gold2009, fraisse2013, planck2014-XXX, bkp2015}. 
To monitor and subtract Galactic foreground contamination many CMB polarimeters observe at multiple 
frequency bands. In this paper we address foreground estimation for CMB polarimeters in the presence of two 
systematic effects: uncertainty in measurement of the frequency band, and frequency dependent polarization 
rotation. 

Band measurement uncertainty is a common systematic effect. Absolute calibration uncertainty, which affects 
all signal components within a given frequency band by the same factor, has already been discussed in 
\citet{stompor2009}, \citet{stivoli2010} and \citet{errard_stompor_2012}. In this paper we focus on band 
measurement uncertainty, which affects sky signals with different emission spectra differently. In the 
context of foreground estimation with unknown foreground spectral parameters, the effect of band measurement 
uncertainties has not yet been studied much in detail. \citet{church03a} gave an analytic calculation of the 
effect of error in the knowledge of bands given sky signals with fixed and known spectral shape. 
\citet{Bao2011} studied the impact of band measurement uncertainty in the presence of an achromatic 
half-wave plate (AHWP) and proposed an approximate dust estimation and subtraction process.

For CMB polarimeters using an AHWP, such as the E and B experiment (EBEX)~\citep{reichborn-kjennerud2010}, 
or using a sinuous antenna multi-chroic pixel~\citep{arnold10}, these elements induce a frequency-dependent 
instrumental polarization rotation. Uncertainties in foreground emission laws and in the relative intensity 
of the CMB and foregrounds - at high frequencies predominantly Galactic dust - induce uncertainties in 
correcting the instrumental polarization rotation. 

The goal of this paper is to develop a foreground estimation formalism in the presence of these two 
systematic effects and to assess its performance in a practical instrumental configuration. Our work is 
based on the maximum likelihood component separation method presented in~\citet{stompor2009}. For 
concreteness we adopt the AHWP model, frequency bands, and approximate noise information that were 
applicable to the design of the EBEX balloon-borne polarimeter. 

In Section~\ref{sec:maxliketheoretical} we explain the theoretical framework of the maximum likelihood 
foreground cleaning algorithm. In Section~\ref{sec:maxlikesimulation} we briefly describe the 
details of our simulation. The results of the simulations are reported in Section~\ref{sec:maxlikeresults}. 
Finally in Section~\ref{sec:maxlikediscussion} we make the concluding remarks.

\section{Theoretical Framework}
\label{sec:maxliketheoretical}

In this section we discuss the mathematical framework of the maximum likelihood foreground estimation.
This framework is developed entirely in the map domain. We start with a brief review of the basic formalism 
in the absence of systematic effects. We then extend it to include band measurement uncertainty and frequency 
dependent polarization rotation.

\subsection{Basic Formalism}
\label{sec:basic_formalism}
We model the sky signal observed in multiple frequency bands for a single pixel in the map as
\begin{equation}
\boldsymbol{d}_p\ =\ \boldsymbol{A}_p\ \boldsymbol{s}_p\ + \ \boldsymbol{n}_p.
\end{equation}
Here the subscript $p$ denotes quantities for a single pixel; \boldmath$d$\unboldmath$_p$~is the data 
vector containing the measured signals for $n_f$ frequency bands and $n_s$ Stokes parameters; 
\boldmath$s$\unboldmath$_p$~is the underlying sky signal vector for $n_c$ sky signal components and $n_s$ 
Stokes parameters; \boldmath$n$\unboldmath$_p$~is the noise vector for $n_f$ frequency bands and $n_s$ 
Stokes parameters; \boldmath$A$\unboldmath$_p \equiv$ \boldmath$A$\unboldmath$_p($\boldmath$\beta$\unboldmath
$)$~is the component `mixing matrix', which maps the sky signals to the observed signals. At each pixel p,
the mixing matrix \boldmath$A$\unboldmath$_p$~has $n_f \times n_c$ blocks. Each block is an $n_s$ by $n_s$ 
diagonal matrix with all of its diagonal elements equal to each other. It is parameterized by a set of 
unknown parameters $\{\beta_i\}$ describing the spectral shape of the components. 
In this paper we assume that the parameters $\{\beta_i\}$ are spatially uniform across the patch. For $n_p$ 
pixels we remove the subscript $p$ and the data model becomes
\begin{equation}
\boldsymbol{d}\ =\ \boldsymbol{A\ s}\ +\ \boldsymbol{n}.
\end{equation}

The likelihood function for the data set is
\begin{equation}
-2\ln \mathcal{L}(\boldsymbol{s, \beta})\ =\ \textrm{const}\ +\ (\boldsymbol{d} - \boldsymbol{A\ s})^{t}
\ \boldsymbol{N}^{-1}\ (\boldsymbol{d} - \boldsymbol{A\ s}),
\label{eq:Ldatadef}
\end{equation}
where \boldmath$N$\unboldmath~is the noise covariance matrix. When there is no correlated noise between 
different pixels, \boldmath$N$\unboldmath~is a rank $n_s \times n_f \times n_p$ square, symmetric, block 
diagonal noise matrix and the full data likelihood can be calculated as the sum of likelihood values 
calculated from each pixel. The likelihood is maximized when
\begin{equation}
\boldsymbol{s}\ =\ (\boldsymbol{A}^{t}\ \boldsymbol{N}^{-1}\ \boldsymbol{A})^{-1}\ \boldsymbol{A}^{t}\ 
\boldsymbol{N}^{-1}\ \boldsymbol{d},\label{eq:smaxlike}
\end{equation}
and by substituting Eq.~\ref{eq:smaxlike} into Eq.~\ref{eq:Ldatadef} we find
\begin{equation}
-2\ln \mathcal{L}\ =\ \textrm{const}\ -\ (\boldsymbol{A}^{t}\ \boldsymbol{N}^{-1}\ 
\boldsymbol{d})^{t}\ (\boldsymbol{A}^{t}\ \boldsymbol{N}^{-1}\ \boldsymbol{A})^{-1}\ 
(\boldsymbol{A}^{t}\ \boldsymbol{N}^{-1}\ \boldsymbol{d}).
\label{eq:Lspecdef}
\end{equation}

For a given set of parameters $\{\beta_i\}$, the mixing matrix \boldmath$A$\unboldmath~can be calculated 
straightforwardly. This suggests a two step foreground estimation process: first finding the set of 
parameters that maximizes the likelihood function (Eq.~\ref{eq:Lspecdef}), then calculating the component 
signals given the maximum likelihood parameters using Eq.~\ref{eq:smaxlike}. 

\subsection{Extension of the Basic Formalism}
\label{sec:extended_formalism}

When extending the basic formalism to include systematic effects, the mixing matrix~\boldmath$A$\unboldmath
~takes a more complicated form while the main steps of the foreground estimation
remain the same.

\subsubsection{Frequency Band Measurement Uncertainty}
\label{sec:addbandeffect}

A frequency band is expressed in terms of transmission as a function of frequency 
$T(\nu)$. We characterize an arbitrary band in terms of two parameters: the band-center and band-width. 
We use the same definitions as in \citet{runyan02a}. The band-center $\nu_c$ is
\begin{equation}
\nu_{c}\ =\ \frac{\int \nu\ T(\nu)\ d\nu}{\int T(\nu)\ d\nu},
\end{equation}
which is the mean frequency weighted by the transmission. The band-width is 
\begin{equation}
\Delta \nu\ =\ \nu_{U}-\nu_{L},
\end{equation}
where the lower and upper band edges are defined as
\begin{equation}
\nu_{U}\ =\ \frac{\int_{\nu_{c}}^{\infty} \nu T^{\prime}(\nu) d\nu}{\int_{\nu_{c}}^{\infty}
          T^{\prime}(\nu) d\nu}\ ,
          \ \nu_{L}\ =\ \frac{\int_{0}^{\nu_{c}} \nu T^{\prime}(\nu) d\nu}{\int_{0}^{\nu_{c}}
          T^{\prime}(\nu) d\nu}, 
\end{equation} 
and $T^{\prime}(\nu)$ is the derivative of the transmission curve $T(\nu)$. This definition of band edges 
favors frequencies where sharp transitions happen in the transmission curve. For example, for a top-hat band 
the lower and upper band edges are where the transmission turns on and off. 

\begin{figure}
\begin{center}
 \scalebox{0.5}{\includegraphics[trim = 10mm 5mm 10mm 10mm, clip]{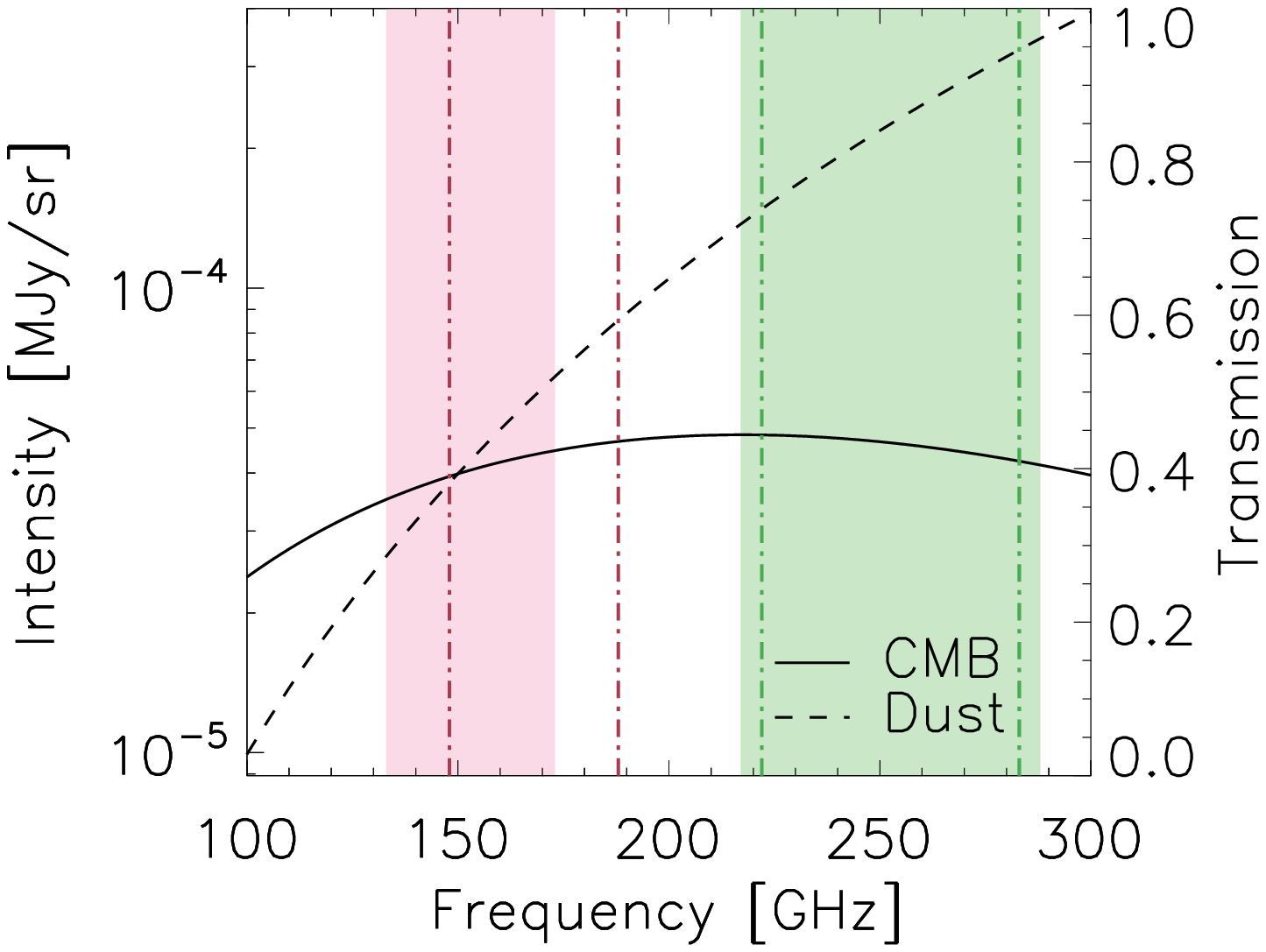}}
 \scalebox{0.5}{\includegraphics[trim = 5mm 5mm 5mm 0mm, clip]{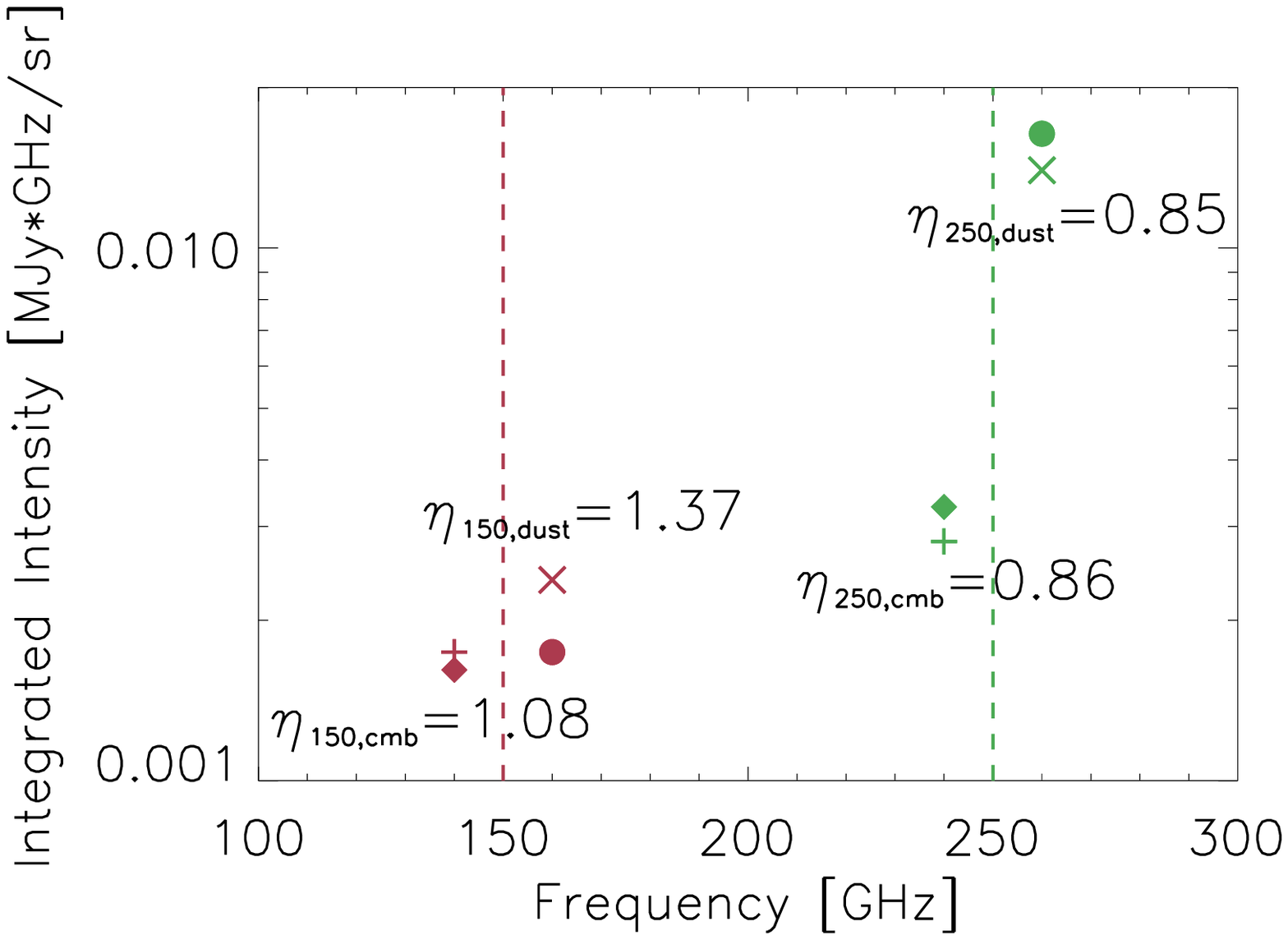}}
 \caption{Illustration of the definition of scaling coefficient \boldmath$\eta$\unboldmath. In this 
   example we have two sky signal components: CMB and polarized Galactic thermal dust emission. 
   There are two top-hat bands centered at 150 and 250 GHz. Left panel: CMB (solid) and Galactic 
   thermal dust (dashed) spectra. For the CMB spectrum, we use the anisotropy spectrum with a 
   0.1~$\mu K_{\textrm{CMB}}$ signal level, which is the inflationary $B$-mode signal level at $\ell = 80$ with
   a tensor-to-scalar ratio $r = 0.1$. The dust spectrum is a 19.6 K blackbody with a 1.59 power law 
   emissivity \citep{planck2014-XXII} scaled to a signal level equal to 0.1~$\mu K_{\textrm{CMB}}$ at 150 GHz. 
   The two solid bands are the assumed bands while the two dot-dashed bands are the nominal bands in the 
   simulation. Compared to the assumed band, the nominal 150 GHz band is shifted by 15 GHz with the same 
   band-width. The nominal 250 GHz band has a 10 GHz less band-width compared to the assumed 250 GHz band 
   while having the same band center. Right panel: the CMB and dust intensity integrated over the bands. The 
   CMB and dust data points are offset along the x-axis for clarity. The scaling coefficients are the ratios 
   of the nominal band-integrated signal (plus - CMB, cross - dust) to the assumed band-integrated signal 
   (diamond - CMB, dot - dust). CMB and dust have different scaling coefficients even with the same band 
   mismatch.}
 \label{fig:eta_visulization}
 \end{center}
 \end{figure}

There are uncertainties in the measurements of frequency bands. To the extent that bands in this paper are 
characterized with two parameters, band center and width, the measured band uncertainties convert to 
uncertainties in these parameters. For any incoming sky signal, band measurement 
uncertainties propagate to uncertainties in the total measured in-band power, which also depends on the 
spectral shape and amplitude of the sky signal. Each sky signal component is affected differently. We 
parameterize the measured power uncertainties in terms of a vector of scaling coefficients 
\boldmath$\eta$\unboldmath~which contains $n_f \times n_c$ components. Here we define the 
\textit{nominal band} as the `true' underlying band of the instrument and the \textit{assumed band} as the 
band determined from the measurements. A scaling coefficient $\eta_{\nu,s}$ is the ratio of the band 
integrated power in the nominal band around frequency $\nu$ and the assumed band for signal component $s$. 
Uncertainties in the band measurements translate to a range of probable $\eta_{\nu,s}$. Because the scaling 
coefficient $\eta_{\nu,s}$ also depends on the emission spectral shape, each sky signal component has a 
different range of probable values. Note that $\eta_{\nu,s}$ does not depend on the overall amplitude of the 
sky components. In the case where the band is measured accurately, the scaling coefficient is unity for 
any sky signal. Figure~\ref{fig:eta_visulization} illustrates the scaling coefficients; it shows 
\boldmath$\eta$\unboldmath~for CMB and polarized Galactic thermal dust in two frequency bands centered at 
150 GHz and 250 GHz. Because in this paper we assume that the spectral shape of the sky components are 
spatially uniform across the patch, the scaling coefficients \boldmath$\eta$\unboldmath~are also spatially 
uniform. 

In our extended foreground estimation formalism we find the maximum likelihood solution for the foreground 
spectral parameters \boldmath$\beta$\unboldmath~{\it and} for the scaling 
coefficients~\boldmath$\eta$\unboldmath. A range of probable scaling coefficients, as obtained from the 
likelihood function, are mappable to possible combinations of band-centers and band-widths. 

The technical implementation is carried out by introducing a new mixing matrix 
\boldmath$A$\unboldmath$_{\textrm{B}}($\boldmath$\beta, \eta$\unboldmath$)$, which has the same rank as the 
original mixing matrix \boldmath$A$\unboldmath, but explicitly includes the scaling coefficients. Each $n_s$ 
by $n_s$ block in \boldmath$A$\unboldmath$_{\textrm{B}}$~is multiplied by the scaling coefficient for the 
corresponding frequency channel and sky component. With \boldmath$A$\unboldmath$_{\textrm{B}}$ the data model 
is
\begin{equation}
\boldsymbol{d} = \boldsymbol{A}_{\textrm{B}}(\boldsymbol{\beta},\boldsymbol{\eta})\ \boldsymbol{s}\ +\ 
\boldsymbol{n}
\end{equation}
and the likelihood function is
\begin{equation}
-2\ln \mathcal{L}_{\textrm{B}}(\boldsymbol{s}, \boldsymbol{\beta}, \boldsymbol{\eta})\ =\ \textrm{const}\ +
\ (\boldsymbol{d} - \boldsymbol{A}_{\textrm{B}}\ \boldsymbol{s})^{t}\ \boldsymbol{N}^{-1}\ 
(\boldsymbol{d} - \boldsymbol{A}_{\textrm{B}}\ \boldsymbol{s}).  
\end{equation}
When the likelihood reaches maximum we have
\begin{equation}
-2\ln \mathcal{L}_{\textrm{B}}\ =\ \textrm{const}\ -\ (\boldsymbol{A}_{\textrm{B}}^{t}\ 
\boldsymbol{N}^{-1}\ \boldsymbol{d})^{t}\ (\boldsymbol{A}_{\textrm{B}}^{t}\ \boldsymbol{N}^{-1}\ 
\boldsymbol{A}_{\textrm{B}})^{-1}\ (\boldsymbol{A}_{\textrm{B}}^{t}\ \boldsymbol{N}^{-1}\ \boldsymbol{d}),
\label{eq:Lspecband}
\end{equation}
and
\begin{equation}
\boldsymbol{s}\ =\ (\boldsymbol{A}_{\textrm{B}}^{t}\ \boldsymbol{N}^{-1}\ \boldsymbol{A}_{\textrm{B}})^{-1}\
\boldsymbol{A}_{\textrm{B}}^{t}\ \boldsymbol{N}^{-1}\ \boldsymbol{d}.
\label{eq:smaxlikeband}
\end{equation}

\begin{figure}
  \begin{center}
    \scalebox{0.7}{\includegraphics{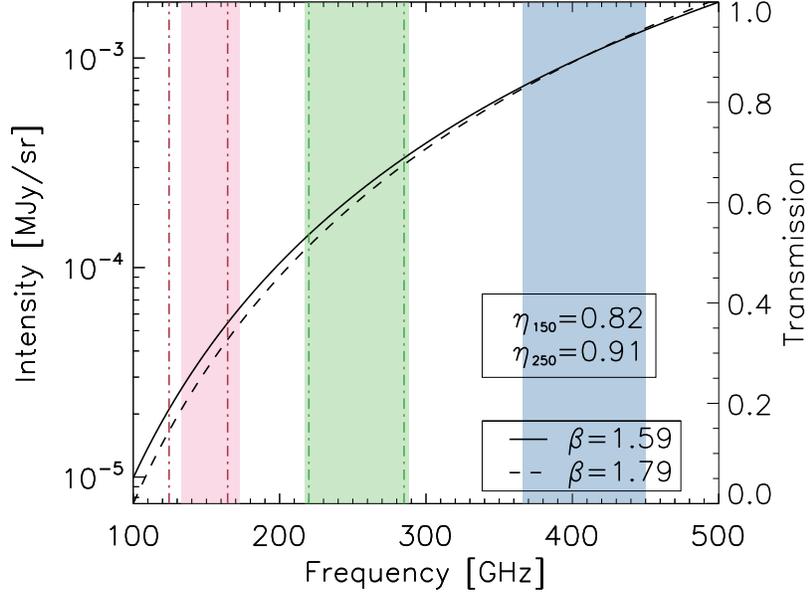}}
    \caption{Illustration of degeneracy between the scaling coefficient for polarized Galactic thermal dust 
      $\eta_{\textrm{d}}$ and the dust spectral index $\beta$. We have an assumed dust spectrum (solid) and 
      three assumed frequency bands centered at 150 GHz, 250 GHz and 410 GHz (filled color bands). The dust 
      spectrum has a spectral index of 1.59 and a signal level of 0.1 $\mu K_{\textrm{CMB}}$ at 150 GHz. If 
      the nominal dust spectrum (dashed) has a spectral index of 1.79 and is normalized to the assumed dust 
      spectrum at 410 GHz band, we have $\eta_{\textrm{150,d}} = 0.82$ and $\eta_{\textrm{250,d}} = 0.91$. If the 
      nominal 150 GHz band is shifted to lower frequency by 8.5 GHz and the nominal 250 GHz band has reduced 
      band-width by 6 GHz compared to the assumed bands, we also have $\eta_{\textrm{150}} = 0.82$ and 
      $\eta_{\textrm{250}} = 0.91$.}
    \label{fig:demoetabetadegeneracy}
  \end{center}
\end{figure}

There exist two degeneracies in the problem. First, the signal level is degenerate with the scaling 
coefficients. For a particular sky component, simultaneously multiplying all the scaling coefficients by a 
factor is equivalent to multiplying the sky signal by the same factor. Second, the spectral parameters and 
the scaling coefficients are degenerate. Tilting the spectrum is equivalent to changing the scaling 
coefficients by corresponding factors. Figure~\ref{fig:demoetabetadegeneracy}~shows an illustration of this 
degeneracy. In the example, a mis-estimate of the polarized Galactic thermal dust spectral index by 0.2 has 
the same $\eta_{\textrm{150,d}}$ and $\eta_{\textrm{250,d}}$ as an 8.5 GHz shift of the band-center at the 150 
band toward lower frequency and a 6 GHz reduction of the band-width at the 250 band.

The degeneracies can be broken by setting priors on the scaling coefficients based on the uncertainty of the 
frequency band measurements. If we assume $\eta_i$ is measured with an uncertainty 
$\sigma_{\eta_i}$ around $\bar{\eta_i}$, the likelihood function becomes
\begin{equation}
-2\ln \mathcal{L}_{\textrm{B}}(\boldsymbol{\beta}, \boldsymbol{\eta})\ =\ \textrm{const}\ -\ 
(\boldsymbol{A}_{\textrm{B}}^{t} \boldsymbol{N}^{-1} \boldsymbol{d})^{t}\ (\boldsymbol{A}_{\textrm{B}}^{t}
\boldsymbol{N}^{-1} \boldsymbol{A}_{\textrm{B}})^{-1}\ (\boldsymbol{A}_{\textrm{B}}^{t}
\boldsymbol{N}^{-1} \boldsymbol{d})\ +\ \sum\limits_i \frac{(\eta_i - \bar{\eta_i})^2}{\sigma_{\eta_i}^2},
\label{eq:Lspecbandwithprior}
\end{equation}
where the last term represents the priors on the scaling coefficients. We use 
Eq.~\ref{eq:Lspecbandwithprior}~to find the best fit spectral parameters and the scaling coefficients. 
We then recover the sky component signals using the best fit parameters.

\subsubsection{Frequency Dependent Polarization Rotation }
\label{sec:addahwpeffect}
Frequency dependent polarization rotation mixes the $Q$ and $U$ signals. The amount of rotation within a 
band depends on the spectral shape of the incoming signals, the characteristic parameters of the instrument, 
and the frequency band shape.

To incorporate instrumental frequency dependent rotation into foreground estimation, we introduce a new 
mixing matrix \boldmath$A$\unboldmath$_{\textrm{R}}($\boldmath$\beta, \theta$\unboldmath$)$~which has the same 
rank as the original mixing matrix \boldmath$A$\unboldmath. The mixing matrix 
\boldmath$A$\unboldmath$_{\textrm{R}}$ is parameterized by the spectral parameters 
\boldmath$\beta$\unboldmath~and a vector \boldmath$\theta$\unboldmath~of rotation angles for each sky signal 
component in each frequency 
band. Compared to \boldmath$A$\unboldmath, the $2 \times 2$ $Q$-$U$ blocks for each sky signal component $s$ 
at each frequency band $\nu$ in \boldmath$A$\unboldmath$_{\textrm{R}}$~are multiplied by rotation matrices 
with non-zero rotation angles \boldmath$\theta$\unboldmath$_{s,\nu}$. Thus each $Q$-$U$ block has non-zero 
off-diagonal elements and \boldmath$A$\unboldmath$_{\textrm{R}}$~is not block diagonal. A requirement on the 
uncertainty of the rotation angles translates to a requirement on the uncertainty of polarization angle
calibration.

With \boldmath$A$\unboldmath$_{\textrm{R}}$~the data model is
\begin{equation}
\boldsymbol{d}\ =\ \boldsymbol{A}_{\textrm{R}}(\boldsymbol{\beta}, \boldsymbol{\theta})\ \boldsymbol{s}\ +\ 
\boldsymbol{n},
\end{equation}
and the likelihood function is
\begin{equation}
-2\ln\ \mathcal{L}_{\textrm{R}}(\boldsymbol{s}, \boldsymbol{\beta}, \boldsymbol{\theta})\ =\ \textrm{const}\ 
+\ (\boldsymbol{d}\ -\ \boldsymbol{A}_{\textrm{R}}\ \boldsymbol{s})^{t}\ \boldsymbol{N}^{-1}\ 
(\boldsymbol{d}\ -\ \boldsymbol{A}_{\textrm{R}}\ \boldsymbol{s}).
\end{equation}
When the likelihood reaches its maximum we have 
\begin{equation}
-2\ln\ \mathcal{L}_{\textrm{R}}\ =\ \textrm{const}\ -\ (\boldsymbol{A}_{\textrm{R}}^{t}\
\boldsymbol{N}^{-1}\ \boldsymbol{d})^{t}\ (\boldsymbol{A}_{\textrm{R}}^{t}\ \boldsymbol{N}^{-1}\
\boldsymbol{A}_{\textrm{R}})^{-1}\ (\boldsymbol{A}_{\textrm{R}}^{t}\ \boldsymbol{N}^{-1}\ \boldsymbol{d}),
\label{eq:Lspecdef_rot}
\end{equation}
and
\begin{equation}
\boldsymbol{s}\ =\ (\boldsymbol{A}_{\textrm{R}}^{t}\ \boldsymbol{N}^{-1}\ \boldsymbol{A}_{\textrm{R}})^{-1}\
\boldsymbol{A}_{\textrm{R}}^{t}\ \boldsymbol{N}^{-1}\ \boldsymbol{d}.
\label{eq:smaxlike_rot}
\end{equation}

There exists a degeneracy between the frequency dependent polarization rotation and the polarization 
angle of the incoming signal. 
Changing the band averaged rotation angles of a sky component in all bands by a given amount is equivalent 
to changing the incoming polarization signal by the same amount. This degeneracy can be broken by setting 
priors on the rotation angles based on the uncertainty of the polarization angle calibration. If we assume 
the band averaged rotation angle $\theta_i$ is determined within an uncertainty $\sigma_{\theta_i}$ around 
$\bar{\theta_i}$ given the knowledge of the instrument and the incoming signals, the likelihood function 
becomes
\begin{equation}
-2\ln \mathcal{L}_{\textrm{R}}(\boldsymbol{\beta}, \boldsymbol{\theta})\ =\ \textrm{const}\ -\ 
(\boldsymbol{A}_{\textrm{R}}^{t}\boldsymbol{N}^{-1}\boldsymbol{d})^{t}\ 
(\boldsymbol{A}_{\textrm{R}}^{t}\boldsymbol{N}^{-1}\boldsymbol{A}_{\textrm{R}})^{-1}\ 
(\boldsymbol{A}_{\textrm{R}}^{t}\boldsymbol{N}^{-1}\boldsymbol{d})\ +\ \sum\limits_i 
\frac{(\theta_i - \bar{\theta_i})^2}{\sigma_{\theta_i}^2},
\label{eq:Lspecrotwithprior}
\end{equation}
where the last term originates from the priors on the rotation angles. 

\subsubsection{Combining Band Measurement Uncertainty and Frequency Dependent Polarization Rotation}
\label{sec:combinebandandahwp}
To incorporate both band measurement uncertainty and frequency dependent polarization rotation at the same 
time we define a new mixing matrix \boldmath$A$\unboldmath$_{\textrm{C}}($\boldmath$\beta$\unboldmath,\boldmath$\eta$\unboldmath,\boldmath$\theta$\unboldmath$)$, which has the same rank as the original mixing
matrix~\boldmath$A$\unboldmath. Compared to~\boldmath$A$\unboldmath, the mixing matrix 
\boldmath$A$\unboldmath$_{\textrm{C}}$ is parameterized by the spectral parameters~\boldmath$\beta$\unboldmath,
the scaling coefficients~\boldmath$\eta$\unboldmath~and the band averaged rotation 
angles~\boldmath$\theta$\unboldmath. Each $n_s$ by $n_s$ block in~\boldmath$A$\unboldmath$_{\textrm{C}}$ is 
multiplied by the in-band scaling coefficient~$\eta_i$ for the corresponding signal component and frequency 
band. The $2 \times 2$ $Q$-$U$ block within each $n_s$ by $n_s$ block is multiplied by a rotation matrix 
with rotation angle $\theta_i$.

With the definition of \boldmath$A$\unboldmath$_{\textrm{C}}$ the data model is
\begin{equation}
\boldsymbol{d}\ =\ \boldsymbol{A}_{\textrm{C}}(\boldsymbol{\beta}, \boldsymbol{\eta}, \boldsymbol{\theta})\ 
\boldsymbol{s}\ +\ \boldsymbol{n},
\end{equation}
and the likelihood function is
\begin{equation}
-2\ln \mathcal{L}_{\textrm{C}}(\boldsymbol{s}, \boldsymbol{\beta}, \boldsymbol{\eta}, \boldsymbol{\theta})\ 
=\ \textrm{const}\ +\ (\boldsymbol{d}\ -\ \boldsymbol{A}_{\textrm{C}}\ \boldsymbol{s})^{t}\ 
\boldsymbol{N}^{-1}\ (\boldsymbol{d}\ -\ \boldsymbol{A}_{\textrm{C}}\ \boldsymbol{s}).
\end{equation}
When the likelihood reaches its maximum we have 
\begin{equation}
-2\ln \mathcal{L}_{\textrm{C}}\ =\ \textrm{const}\ -\ (\boldsymbol{A}_{\textrm{C}}^{t}\ 
\boldsymbol{N}^{-1}\ \boldsymbol{d})^{t}\ 
(\boldsymbol{A}_{\textrm{C}}^{t}\ \boldsymbol{N}^{-1}\ \boldsymbol{A}_{\textrm{C}})^{-1}\ 
(\boldsymbol{A}_{\textrm{C}}^{t}\ \boldsymbol{N}^{-1}\ \boldsymbol{d}),
\label{eq:Lspecdef_combined}
\end{equation}
and
\begin{equation}
\boldsymbol{s}\ =\ (\boldsymbol{A}_{\textrm{C}}^{t}\ \boldsymbol{N}^{-1}
\ \boldsymbol{A}_{\textrm{C}})^{-1}\ \boldsymbol{A}_{\textrm{C}}^{t}\ 
\boldsymbol{N}^{-1}\ \boldsymbol{d}.
\label{eq:smaxlike_combined}
\end{equation}

Due to the degeneracies mentioned in Sects.~\ref{sec:addbandeffect} and~\ref{sec:addahwpeffect}, to 
recover the signals accurately we set priors to the scaling coefficients~\boldmath$\eta$\unboldmath~and the 
band averaged rotation angles~\boldmath$\theta$\unboldmath~based on the uncertainty of the measurements. 
The likelihood function becomes
\begin{multline}
-2\ln \mathcal{L}_{\textrm{C}}(\boldsymbol{\beta}, \boldsymbol{\eta}, \boldsymbol{\theta})\ =\ 
\textrm{const}\ -\ (\boldsymbol{A}_{\textrm{C}}^{t}\ \boldsymbol{N}^{-1}\
\boldsymbol{d})^{t}\ (\boldsymbol{A}_{\textrm{C}}^{t}\ 
\boldsymbol{N}^{-1}\ \boldsymbol{A}_{\textrm{C}})^{-1}\ (\boldsymbol{A}_{\textrm{C}}^{t}\
\boldsymbol{N}^{-1}\ \boldsymbol{d})\\
 +\ \sum\limits_i \frac{(\eta_i - \bar{\eta_i})^2}{\sigma_{\eta_i}^2}\ +
\ \sum\limits_i \frac{(\theta_i - \bar{\theta_i})^2}{\sigma_{\theta_i}^2},
\label{eq:Lspeccombinedwithprior}
\end{multline}
where the last two terms come from the priors.

\subsection{Error Propagation}
\label{sec:errorprop}
Uncertainties in the spectral parameters and instrumental parameters in the mixing matrix propagate to the 
estimated sky component signals. We estimate the covariance matrix for the parameters using the inverse of 
the curvature matrix calculated at the maximum likelihood. In the basic case without systematic effects we 
have
\begin{equation}
\mbox{\boldmath$\tilde{N}$\unboldmath}_{\beta\beta}\ =\ [
(\boldsymbol{A},_{\beta_i}\boldsymbol{s})^t\ \boldsymbol{N}^{-1}\
(\boldsymbol{A},_{\beta_j}\boldsymbol{s}) - 
(\boldsymbol{A},_{\beta_i\beta_j}\boldsymbol{s})^t\
\boldsymbol{N}^{-1}\ (\boldsymbol{d} - \boldsymbol{A\ s}) - \boldsymbol{M}^t \boldsymbol{\hat{N} M}]^{-1}
\label{eqn:Nbetabetadef}
\end{equation}
for the spectral parameters \boldmath$\beta$\unboldmath, where \boldmath$\hat{N}$\unboldmath~is the 
estimated uncertainty matrix of the sky component maps assuming the mixing matrix is perfectly known 
\begin{equation}
\boldsymbol{\hat{N}}\ =\ (\boldsymbol{A}^t\ \boldsymbol{N}^{-1}\ \boldsymbol{A})^{-1},
\end{equation}
and the \boldmath$M$\unboldmath~matrix is defined as
\begin{equation}
\boldsymbol{M}\ =\ \boldsymbol{A}^t\ \boldsymbol{N}^{-1}\ \boldsymbol{A},_{\beta}\boldsymbol{s} - 
\boldsymbol{A}^t,_{\beta}\ \boldsymbol{N}^{-1}\ (\boldsymbol{d} - \boldsymbol{A\ s}).
\label{eqn:Mmatrixdef}
\end{equation}
\boldmath$A,_{\beta}$\unboldmath~and \boldmath$A,_{\beta\beta}$\unboldmath~are the first and second order 
derivatives of the mixing matrix with respect to the spectral parameters, estimated at the best fit values 
of the spectral parameters. The noise matrix of the estimated sky signals is
\begin{equation}
\boldsymbol{\tilde{N}}_{ss} = \boldsymbol{\hat{N}} + (\boldsymbol{\hat{N}M})\ 
\boldsymbol{\tilde{N}}_{\beta\beta}\ (\boldsymbol{\hat{N}M})^t.
\label{eqn:Nssdef}
\end{equation}

When systematic effects are considered, the mixing matrix \boldmath$A$\unboldmath~in 
Eq.~\ref{eqn:Nbetabetadef}, Eq.~\ref{eqn:Mmatrixdef} and Eq.~\ref{eqn:Nssdef}, is changed to the 
corresponding extended mixing matrix (\boldmath$A$\unboldmath$_{\textrm{B}}$, 
\boldmath$A$\unboldmath$_{\textrm{R}}$~or \boldmath$A$\unboldmath$_{\textrm{C}}$) and the derivative of the 
mixing matrix is performed with respect to all corresponding unknown parameters including the spectral 
parameters \boldmath$\beta$\unboldmath, the scaling coefficients \boldmath$\eta$\unboldmath~and the band 
averaged rotation angles \boldmath$\theta$\unboldmath. 

\section{Simulations}
\label{sec:maxlikesimulation}
\begin{figure}
\begin{center}
 \scalebox{0.43}{\includegraphics[trim = 10mm 0mm 35mm 10mm, clip]{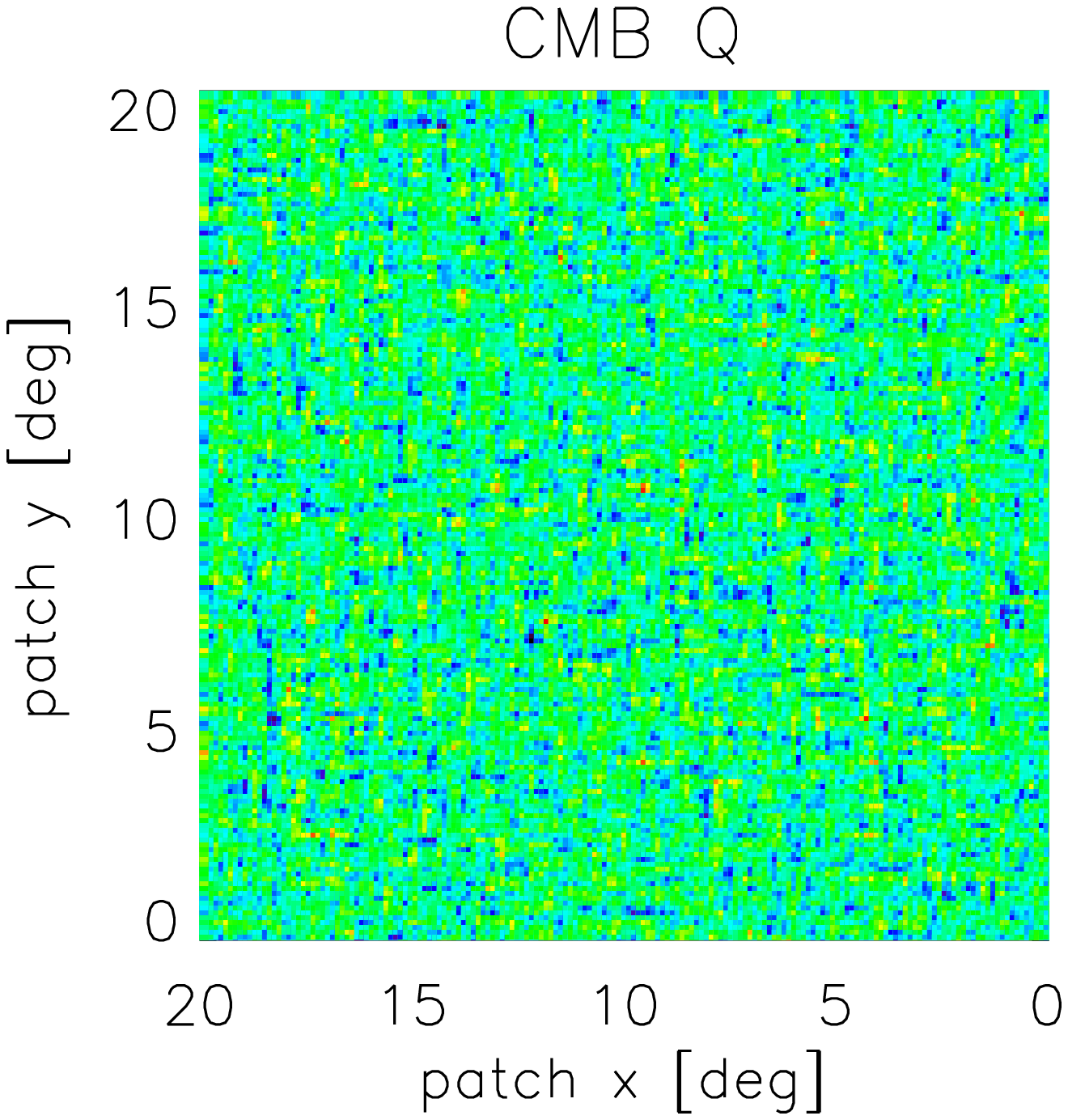}}
 \scalebox{0.43}{\includegraphics[trim = 25mm 0mm 10mm 10mm, clip]{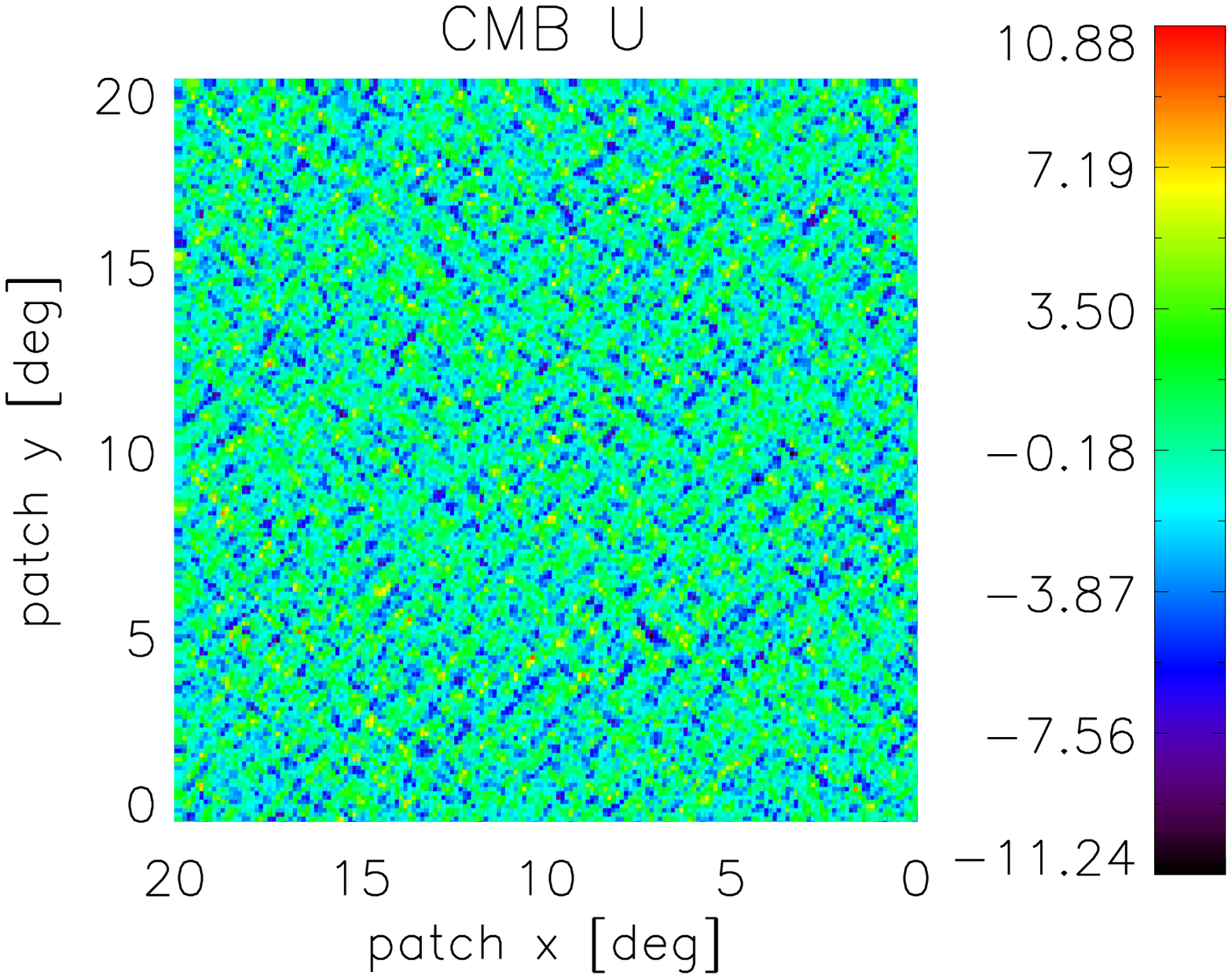}}
 \scalebox{0.43}{\includegraphics[trim = 10mm 0mm 35mm 0mm, clip]{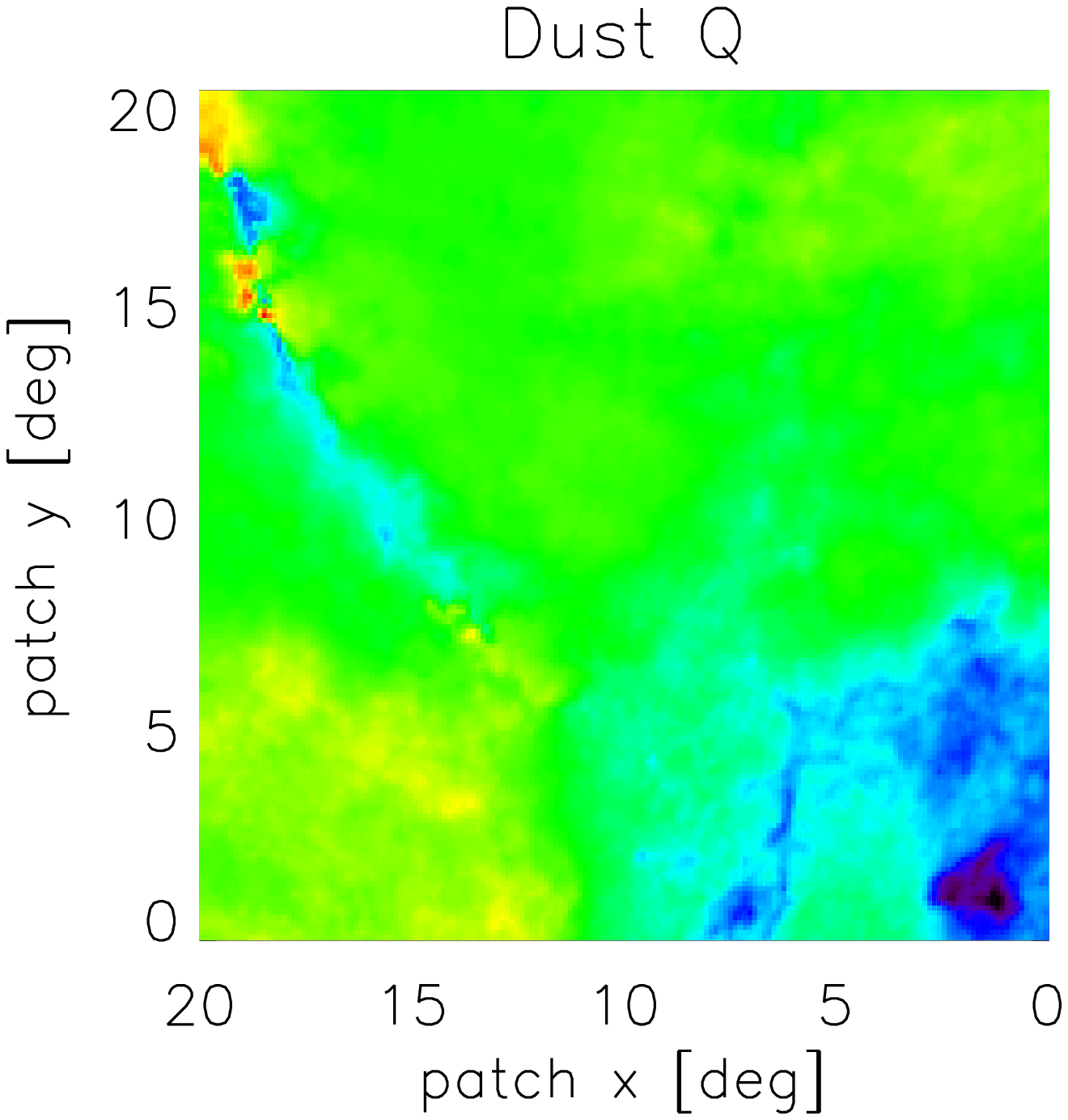}}
 \scalebox{0.43}{\includegraphics[trim = 25mm 0mm 10mm 0mm, clip]{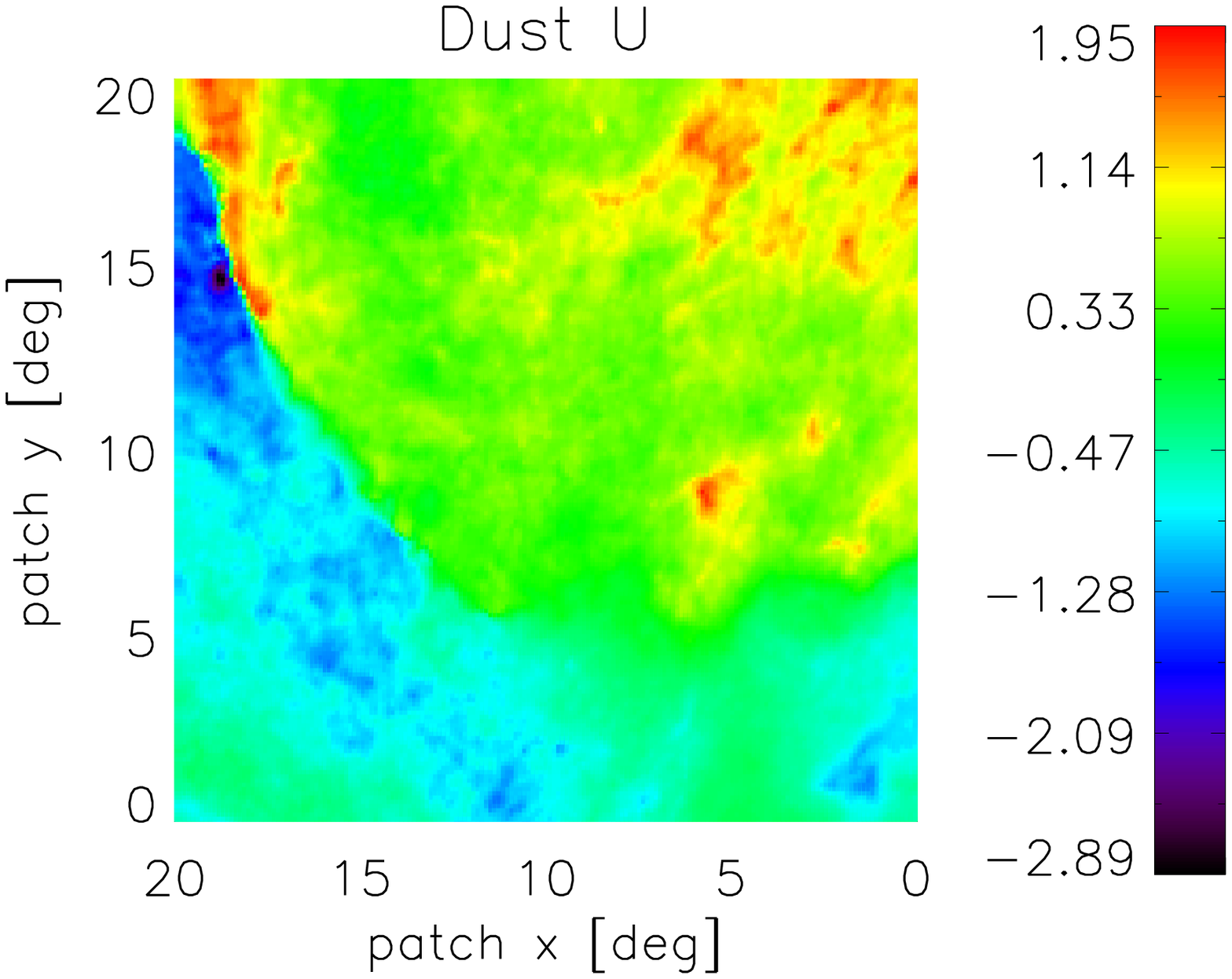}}
 \caption{$Q$ (left column) and $U$ (right column) maps of one realization of CMB (top row) and dust 
   (bottom row) at 150 GHz band on the sky patch chosen for simulation. The maps are in units of 
   $\mu K_{\textrm{RJ}}$. The dust signal is simulated using the same procedure as \citet{stivoli2010}.
   The polarization fraction is assumed to be 10\%, while the polarization angle is derived by the WMAP 
   observations on large scales ($\ell \lesssim 100$), and Gaussian power at small scales.}
 \label{fig:patch_cmb_dust_maps}
 \end{center}
 \end{figure}

We test the extended foreground estimation formalism using simulations. We select a 
$20^{\circ} \times 20^{\circ}$ sky patch centered on RA = $55^{\circ}$, DEC = $-45^{\circ}$ and include only the 
CMB and Galactic dust as sky signals. The input CMB angular power spectra are generated with Code 
for Anisotropies in the Microwave Background (CAMB) described in \citet{lewis00} using the Wilkinson 
Microwave Anisotropy Probe (WMAP) seven-year best-fit cosmological parameters \citep{komatsu11} and a 
tensor to scalar ratio of $r = 0.05$. 
With a flat-sky approximation \citep{kaiser92} we generate random Fourier amplitudes in temperature 
and polarization to match the $C_\ell$ power spectra, which are then transformed to CMB maps on the chosen 
square patch of sky. We adopt the process to generate Galactic dust foreground detailed in \citet{stivoli2010}. 
The dust intensity and its frequency scaling are given by an $19.6$~K blackbody with a 1.59 power-law 
emissivity according to the recent Planck measurement \citep{planck2014-XXII}. The dust polarization fraction 
is set to 10\%. Both the dust frequency scaling and the dust polarization fraction are assumed to be 
spatially uniform. The large scale polarization angle patterns ($\ell \lesssim 100$) are derived from WMAP 
dust polarization template \citep{page2007}. On small angular scales a Gaussian fluctuation power is added 
using the recipe described in \citet{giardino02}. Figure~\ref{fig:patch_cmb_dust_maps}~shows the $Q$ and $U$ 
maps of the dust signal and one realization of CMB in the selected sky patch.

The observations are simulated in three top-hat bands centered at 150~GHz, 250~GHz and 410~GHz. The 
frequency ranges for the three bands and their noise levels per $6.87^{\prime} \times 6.87^{\prime}$ square 
pixel in the $Q$ and $U$ maps are listed in Table \ref{tab:ebexbandsandnoise}.  The noise is assumed to be 
uniform and white across the patch. The noise realization is added to the signal in the map domain in all 
simulations.
\begin{table}
\begin{center}
\begin{tabular}{|c|c|c|}
\hline
Top-hat band & Frequency range [GHz] & Pixel Noise [$\mu K_{\textrm{RJ}}$] \\ \hline
150 GHz & [133, 173] & 0.8 \\ \hline
250 GHz & [217, 288] & 1.0 \\ \hline
410 GHz & [366, 450] & 1.4 \\ \hline
\end{tabular}
\caption{Top-hat bands and the corresponding noise per $6.87^{\prime} \times 6.87^{\prime}$ pixel in the $Q$ 
  and $U$ maps used in the simulation.}
\label{tab:ebexbandsandnoise}
\end{center}
\end{table}

\begin{figure}[h]
\begin{center}
 \scalebox{0.43}{\includegraphics[trim = 10mm 0mm 35mm 10mm, clip]{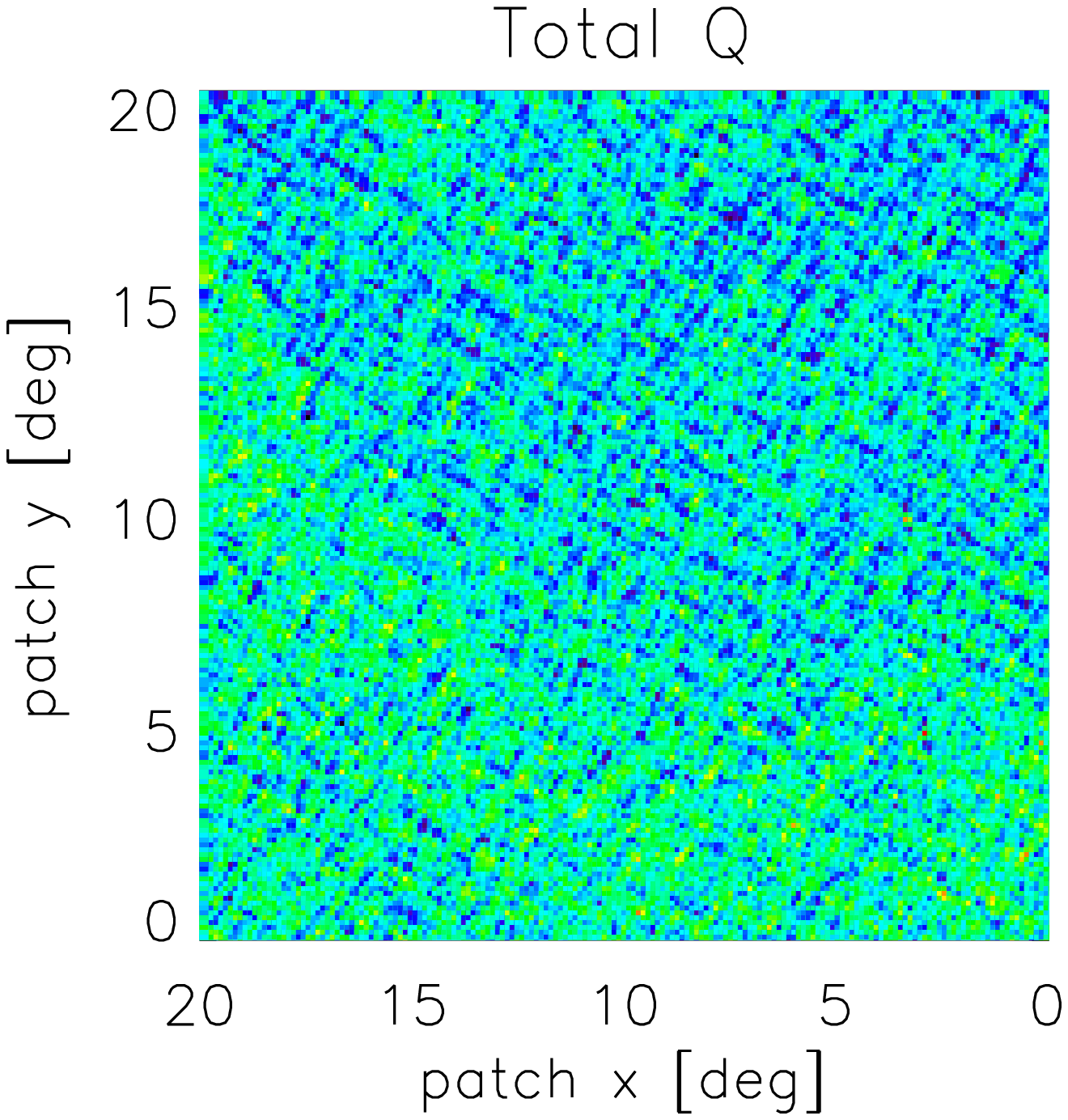}}
 \scalebox{0.43}{\includegraphics[trim = 25mm 0mm 10mm 10mm, clip]{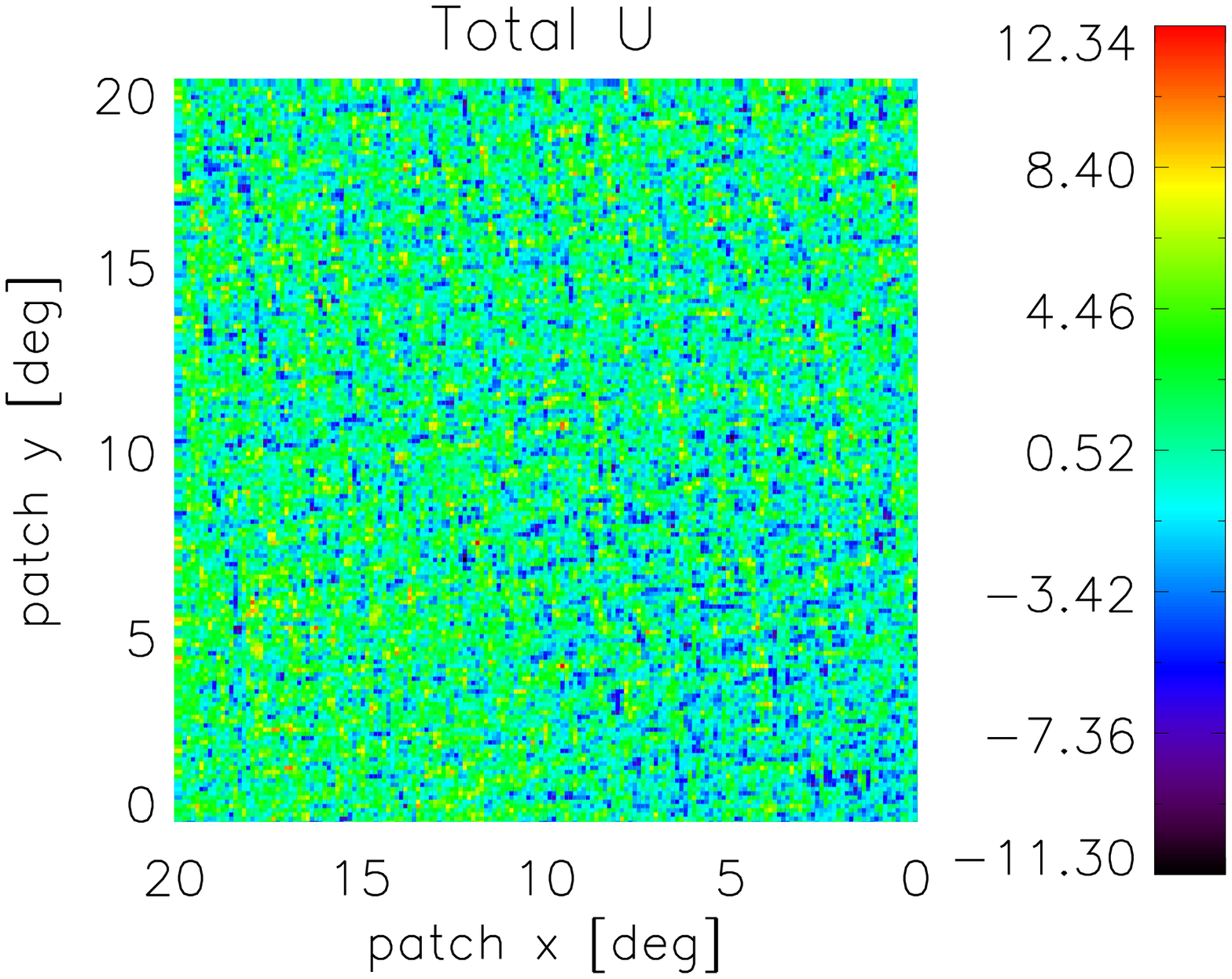}}
 \caption{$Q$ (left) and $U$ (right) maps of observation at 150 GHz band with AHWP rotation effect and 
   instrumental noise included. This is one realization of the CMB and instrumental noise. The maps are in 
   units of $\mu K_{\textrm{RJ}}$.}
 \label{fig:patch_total_rotated_maps}
 \end{center}
 \end{figure}

In the simulation we include a continuously rotating achromatic half-wave plate (AHWP) which induces a 
frequency dependent polarization rotation. The AHWP is composed of a stack of five single sapphire 
half-wave plates with an average thickness of 1.65 mm. The relative orientation angles between the optical 
axis of the plates and the first plate in the stack are 0$^\circ$, 28$^\circ$, 94$^\circ$, 28$^\circ$, 
0$^\circ$. The ordinary and extraordinary refractive indices of sapphire at cryogenic temperature are used 
to calculate the instrumental polarization rotation angles, i.e., $n_o\ =\ 3.047$ and 
$ n_e\ =\ 3.361$~\citep{afsar91,Loewenstein73}. We model the AHWP using Mueller matrices formalism
described in \citet{matsumura2009}. Given the frequency bands, the input dust 
spectrum and the AHWP parameters, the rotation angles are 115.66$^{\circ}$, 103.64$^{\circ}$, 110.76$^{\circ}$ 
for CMB and 115.02$^{\circ}$, 103.68$^{\circ}$, 112.88$^{\circ}$ for dust at 150, 250, 410 GHz bands, 
respectively . Figure~\ref{fig:patch_total_rotated_maps}~shows the observed $Q$ and $U$ maps at 150 GHz band. 
They include both CMB, and dust, rotation induced by the AHWP, and instrumental noise. The parameters of the 
AHWP and the noise levels were chosen to be close to those designed for the EBEX balloon-borne instrument.

The $EE$ and $BB$ power spectra are calculated simultaneously using a flat-sky approximation 
\citep{kaiser92}. For each set of input parameters, the simulation is run 100 times with different CMB and 
noise realizations while the dust signal is kept the same. The result shown for any given $\ell$ bin is the 
mean of the 100 simulations and the error bar is the standard deviation. Specifically, the error 
bars of the input CMB spectrum are the cosmic variances while the error bars of the output CMB spectrum also
include uncertainties from the instrumental noise and the propagation of the parameter uncertainties. We 
also calculate residual power spectra. For each of the 100 simulations the residual spectrum is the 
difference between the estimated CMB spectrum, after estimation of the foreground, and the input spectrum. 
We plot the mean and the standard deviations of the 100 simulations. We consider the estimated CMB $B$-mode 
power spectrum `biased' using two criteria such that each provides somewhat different information: (1) when 
the estimated CMB $B$-mode signal power is more than twice the cosmic variance away from the input in any 
$\ell$ bin, and (2) when the residual power spectrum is more than one sigma away from zero. The first 
criterion quantifies bias relative to irreducible cosmic variance and is particularly useful for noiseless 
simulations. The second is more relevant when instrument noise is included and is less stringent than the 
first.

For the simulations the data model in the basic formalism
\begin{equation}
\boldsymbol{d\ =\ A\ s\ +\ n},
\end{equation}
becomes
\begin{equation}
\scalemath{0.78}{\left(\begin{array}{c}
  I_{150}\\
  Q_{150}\\
  U_{150}\\
  I_{250}\\
  Q_{250}\\
  U_{250}\\
  I_{410}\\
  Q_{410}\\
  U_{410}
\end{array}\right)
 = 
\left(\begin{array}{cccccc}
  A_{150,{\textrm{c}}} & 0 & 0 & A_{150,{\textrm{d}}}(\beta) & 0 & 0 \\
  0 & A_{150,{\textrm{c}}} & 0 & 0 & A_{150,{\textrm{d}}}(\beta) & 0 \\
  0 & 0 & A_{150,{\textrm{c}}} & 0 & 0 & A_{150,{\textrm{d}}}(\beta) \\
  A_{250,{\textrm{c}}} & 0 & 0 & A_{250,{\textrm{d}}}(\beta) & 0 & 0 \\
  0 & A_{250,{\textrm{c}}} & 0 & 0 & A_{250,{\textrm{d}}}(\beta) & 0 \\
  0 & 0 & A_{250,{\textrm{c}}} & 0 & 0 & A_{250,{\textrm{d}}}(\beta) \\
  A_{410,{\textrm{c}}} & 0 & 0 & A_{410,{\textrm{d}}}(\beta) & 0 & 0 \\
  0 & A_{410,{\textrm{c}}} & 0 & 0 & A_{410,{\textrm{d}}}(\beta) & 0 \\
  0 & 0 & A_{410,{\textrm{c}}} & 0 & 0 & A_{410,{\textrm{d}}}(\beta) 
\end{array}\right)
\left(\begin{array}{c}
  I_{CMB}\\
  Q_{CMB}\\
  U_{CMB}\\
  I_{dust}\\
  Q_{dust}\\
  U_{dust}
\end{array}\right)
+ \left(\begin{array}{c}
  n_{I_{150}}\\
  n_{Q_{150}}\\
  n_{U_{150}}\\
  n_{I_{250}}\\
  n_{Q_{250}}\\
  n_{U_{250}}\\
  n_{I_{410}}\\
  n_{Q_{410}}\\
  n_{U_{410}}
\end{array}\right)}.
\end{equation}
Here $I$, $Q$ and $U$ are the polarization stokes parameters. The subscripts denote frequency bands and 
components. The elements in the mixing matrix related to CMB are known since we know the CMB spectrum. For 
any given frequency $\nu$, the element in the mixing matrix related to dust is
\begin{equation}
A_{\nu,{\textrm{d}}}(\beta)\ =\ \Big(\frac{\nu}{\nu_0}\Big)^{\beta}B_{\nu}(T_{\textrm{d}},\nu),
\end{equation}
where $\nu_0$ is a reference frequency typically set to the center of the highest frequency channel of the 
experiment, $B_{\nu}$ is the blackbody spectrum, $T_{\textrm{d}}$ is the dust temperature (which is assumed to 
be known) and $\beta$ is the spectral index of dust which is the sole unknown parameter in this case.

When band measurement uncertainties are included, the mixing matrix becomes
\begin{equation}
\boldsymbol{A}_{\textrm{B}}
 = \left(
\scalemath{0.78}{
\begin{array}{cccccc}
  A_{150,{\textrm{c}}} \eta_{150,{\textrm{c}}} & 0 & 0 & A_{150,{\textrm{d}}}(\beta) \eta_{150,{\textrm{d}}} & 0 & 0 \\
  0 & A_{150,{\textrm{c}}} \eta_{150,{\textrm{c}}} & 0 & 0 & A_{150,{\textrm{d}}}(\beta) \eta_{150,{\textrm{d}}} & 0 \\
  0 & 0 & A_{150,{\textrm{c}}} \eta_{150,{\textrm{c}}} & 0 & 0 & A_{150,{\textrm{d}}}(\beta) \eta_{150,{\textrm{d}}} \\
  A_{250,{\textrm{c}}} \eta_{250,{\textrm{c}}} & 0 & 0 & A_{250,{\textrm{d}}}(\beta) \eta_{250,{\textrm{d}}} & 0 & 0 \\
  0 & A_{250,{\textrm{c}}} \eta_{250,{\textrm{c}}} & 0 & 0 & A_{250,{\textrm{d}}}(\beta) \eta_{250,{\textrm{d}}} & 0 \\
  0 & 0 & A_{250,{\textrm{c}}} \eta_{250,{\textrm{c}}} & 0 & 0 & A_{250,{\textrm{d}}}(\beta) \eta_{250,{\textrm{d}}} \\
  A_{410,{\textrm{c}}} \eta_{410,{\textrm{c}}} & 0 & 0 & A_{410,{\textrm{d}}}(\beta) \eta_{410,{\textrm{d}}} & 0 & 0 \\
  0 & A_{410,{\textrm{c}}} \eta_{410,{\textrm{c}}} & 0 & 0 & A_{410,{\textrm{d}}}(\beta) \eta_{410,{\textrm{d}}} & 0 \\
  0 & 0 & A_{410,{\textrm{c}}} \eta_{410,{\textrm{c}}} & 0 & 0 & A_{410,{\textrm{d}}}(\beta) \eta_{410,{\textrm{d}}} \\
\end{array}
}\right).
\end{equation}
There are seven unknown parameters in \boldmath$A$\unboldmath$_{\textrm{B}}$: the dust spectral index and six 
scaling coefficients.

When the frequency dependent polarization rotation is included, the mixing matrix becomes
\begin{equation}
\boldsymbol{A}_{\textrm{R}}
 = \left(
\begin{array}{cc}
 \boldsymbol{A}_{\textrm{$R_150,\textrm{c}$}} & \boldsymbol{A}_{\textrm{$R_{150,d}$}}\\
 \boldsymbol{A}_{\textrm{$R_{250,c}$}} & \boldsymbol{A}_{\textrm{$R_{250,d}$}}\\
 \boldsymbol{A}_{\textrm{$R_{410,c}$}} & \boldsymbol{A}_{\textrm{$R_{410,d}$}}\\
\end{array}
\right),
\end{equation}
with the blocks \boldmath$A$\unboldmath$_{\textrm{$R_{\nu,s}$}}$~for CMB and dust at frequency band $\nu$ being
\begin{displaymath}
 \boldsymbol{A}_{\textrm{$R_{\nu,c}$}}
 = \left(
\begin{array}{ccc}
  A_{\nu,{\textrm{c}}} & 0 & 0 \\
  0& A_{\nu,{\textrm{c}}} \cos \theta_{\nu,{\textrm{c}}} & -A_{\nu,{\textrm{c}}} \sin \theta_{\nu,{\textrm{c}}}\\
  0& A_{\nu,{\textrm{c}}} \sin \theta_{\nu,{\textrm{c}}} & A_{\nu,{\textrm{c}}} \cos \theta_{\nu,{\textrm{c}}} \\
\end{array}
\right),
\end{displaymath}
\begin{displaymath}
 \boldsymbol{A}_{\textrm{$R_{\nu,d}$}}
 = \left(
\begin{array}{ccc}
A_{\nu,{\textrm{d}}}(\beta) & 0 & 0 \\
0 & A_{\nu,{\textrm{d}}}(\beta) \cos \theta_{\nu,{\textrm{d}}}(\beta)& -A_{\nu,{\textrm{d}}}(\beta) \sin 
\theta_{\nu,{\textrm{d}}}(\beta) \\
0 & A_{\nu,{\textrm{d}}}(\beta) \sin \theta_{\nu,{\textrm{d}}}(\beta) & A_{\nu,{\textrm{d}}}(\beta) \cos 
\theta_{\nu,{\textrm{d}}}(\beta) \\
\end{array}
\right),
\end{displaymath}
where the subscript $\nu$ runs over 150, 250, and 410 GHz.
There are seven unknown parameters in \boldmath$A$\unboldmath$_{\textrm{R}}$: the dust spectral index and six 
rotation angles. 

When both the band measurement uncertainty and the frequency dependent rotation effect of the AHWP are 
considered, the mixing matrix becomes
\begin{equation}
\boldsymbol{A}_{\textrm{C}}
 = \left(
\begin{array}{cc}
 \boldsymbol{A}_{\textrm{$C_{150,c}$}} & \boldsymbol{A}_{\textrm{$C_{150,d}$}}\\
 \boldsymbol{A}_{\textrm{$C_{250,c}$}} & \boldsymbol{A}_{\textrm{$C_{250,d}$}}\\
 \boldsymbol{A}_{\textrm{$C_{410,c}$}} & \boldsymbol{A}_{\textrm{$C_{410,d}$}}\\
\end{array}
\right),
\end{equation}
with the blocks \boldmath$A$\unboldmath$_{\textrm{C}_{\nu,s}}$~for CMB and dust at frequency channel $\nu$ being
\begin{displaymath}
 \boldsymbol{A}_{\textrm{C}_{\nu,\textrm{c}}}
 = \left(
\begin{array}{ccc}
  \eta_{\nu,{\textrm{c}}} A_{\nu,{\textrm{c}}} & 0 & 0 \\
  0& \eta_{\nu,{\textrm{c}}} A_{\nu,{\textrm{c}}} \cos \theta_{\nu,{\textrm{c}}} & -\eta_{\nu,{\textrm{c}}} A_{\nu,{\textrm{c}}} \sin \theta_{\nu,{\textrm{c}}} \\
  0& \eta_{\nu,{\textrm{c}}} A_{\nu,{\textrm{c}}} \sin \theta_{\nu,{\textrm{c}}} & \eta_{\nu,{\textrm{c}}} A_{\nu,{\textrm{c}}} \cos \theta_{\nu,{\textrm{c}}} \\
\end{array}
\right),
\end{displaymath}
\begin{displaymath}
 \boldsymbol{A}_{\textrm{C}_{\nu,\textrm{d}}}
 = \left(
\begin{array}{ccc}
  \eta_{\nu,{\textrm{d}}} A_{\nu,{\textrm{d}}}(\beta) & 0 & 0 \\
  0 & \eta_{\nu,{\textrm{d}}} A_{\nu,{\textrm{d}}}(\beta) \cos \theta_{\nu,{\textrm{d}}}(\beta)& -\eta_{\nu,{\textrm{d}}} A_{\nu,{\textrm{d}}}(\beta) \sin \theta_{\nu,{\textrm{d}}}(\beta) \\
  0 & \eta_{\nu,{\textrm{d}}} A_{\nu,{\textrm{d}}}(\beta) \sin \theta_{\nu,{\textrm{d}}}(\beta) & \eta_{\nu,{\textrm{d}}} A_{\nu,{\textrm{d}}}(\beta) \cos \theta_{\nu,{\textrm{d}}}(\beta) \\
\end{array}
\right),
\end{displaymath}
where the subscript $\nu$ runs over 150, 250 and 410 GHz.
There are a total of 13 unknown parameters in \boldmath$A$\unboldmath$_{\textrm{C}}$: the dust spectral index, 
six scaling coefficients and six band averaged rotation angles. Note that 
\boldmath$A$\unboldmath$_{\textrm{C}}$ reduce to \boldmath$A$\unboldmath$_{\textrm{B}}$ in the 
case of $\theta_{\nu,{\textrm{c}}} = \theta_{\nu,{\textrm{d}}} = 0$ and reduce to 
\boldmath$A$\unboldmath$_{\textrm{R}}$ in the case of $\eta_{\nu,{\textrm{c}}} = \eta_{\nu,{\textrm{d}}} = 1$.
In our simulations, when setting priors to parameters, we assume a Gaussian prior with quoted FWHM values.

\section{Results}
\label{sec:maxlikeresults}
\subsection{Consistency Check}

\begin{figure}[!h]
  \begin{center}
    \scalebox{0.5}{\includegraphics{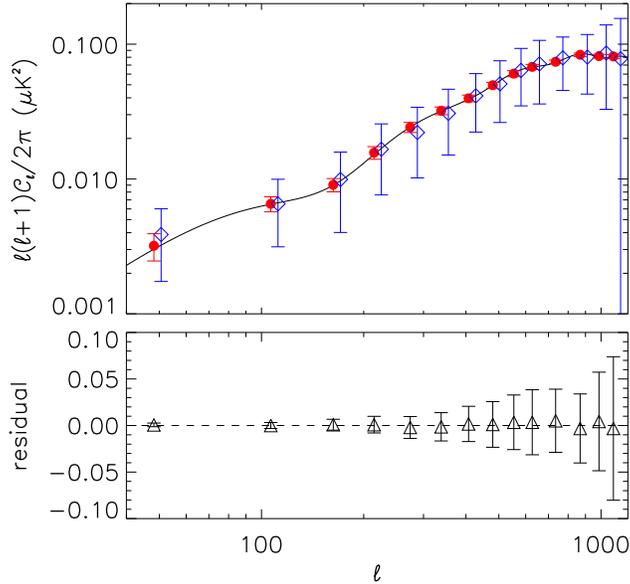}}
  \caption{Demonstration of the validity of the extended formalism assuming perfect knowledge of the 
      scaling coefficients and the rotation angles. The dust spectral index is the only parameter 
      estimated here. Top: $B$-mode angular power spectra of the input CMB~(red dot, error bars 
      include only cosmic variance) and the estimated CMB~(blue diamond). No bias is introduced in the 
      final estimated CMB $B$-mode signal compared to the input. The solid black line is the theoretical 
      CMB $B$-mode power spectrum assuming $r = 0.05$. Bottom: residual angular power spectrum between 
      the estimated CMB $B$-mode signal and the input. The residual (black triangle) is consistent with 
      zero (dashed line) in all $\ell$ bins.}
    \label{fig:nom_noise_fitbetaonly}
  \end{center}
\end{figure}

In the absence of band measurement and polarization angle calibration uncertainties the extended formalism 
reproduces the input spectra with no bias; this is shown in Fig.~\ref{fig:nom_noise_fitbetaonly}. For this 
simulation we {\it included} the systematic effects but they were assumed to be perfectly known. In all the 
power spectra plots shown in the rest of this section, the data points of the estimated CMB $B$-mode power 
spectrum are plotted offset along the x-axis for clarity. We also plot the theoretical underlying CMB 
$B$-mode power spectrum with $r = 0.05$. 

\begin{figure}[h]
  \plottwo{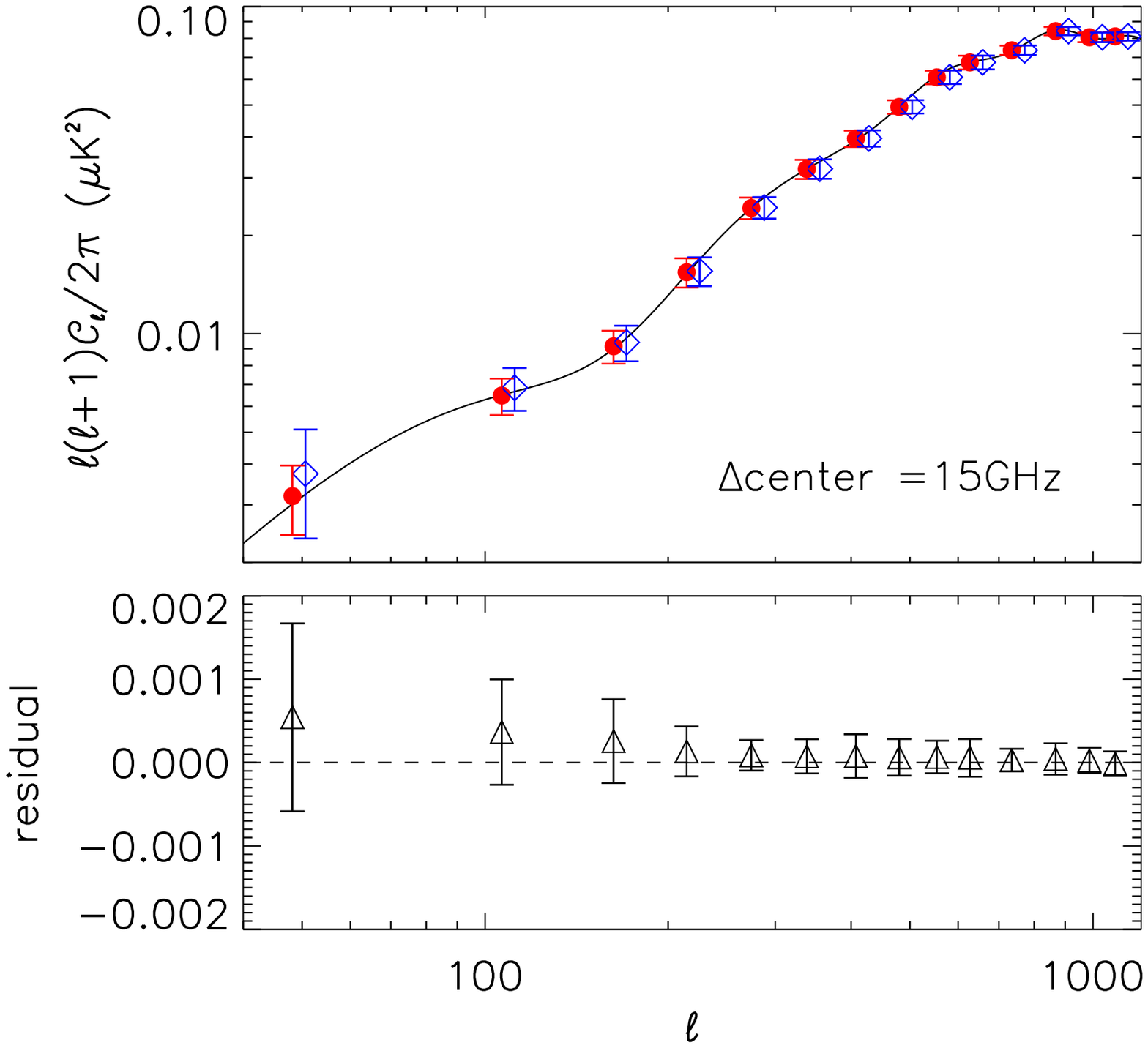}{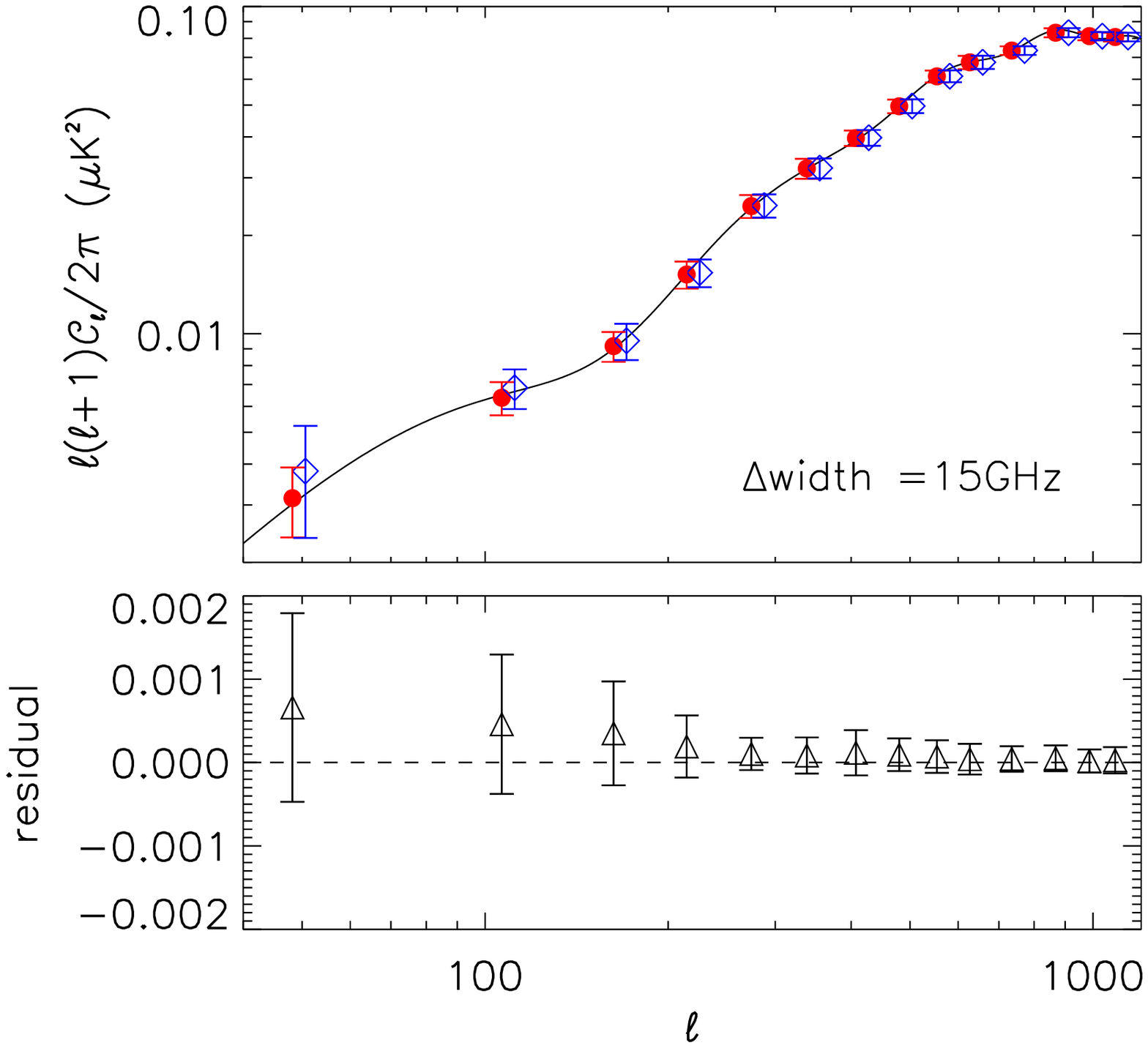}
    \caption{Simulation results where the band measurement uncertainty and frequency dependent polarization
      rotation are included only for 150 GHz band in the absence of instrumental noise. The 250 GHz 
      and 410 GHz bands are assumed to be perfectly known. The band-center (left) or the band-width (right) 
      of the 150~GHz band is mis-estimated by 15~GHz. In both cases the estimated the CMB $B$-mode signal 
      (blue diamond) is not biased compared to the input signal (red dot), and the residual (black triangle) 
      is consistent with zero.}
    \label{fig:combine150s15}
\end{figure}

Next we assess the intrinsic bias in the estimated CMB signal induced by the formalism in the absence of 
instrumental noise. Due to the degeneracies discussed in Sec.~\ref{sec:extended_formalism}, here we consider
the systematic effects in only one band at a time while assuming the other two bands are perfectly known. 
All the parameters in the mixing matrix are estimated without any priors. We test the formalism including 
band measurement uncertainty only, frequency dependent polarization rotation only, or both systematic 
effects. The estimated CMB $B$-mode signal is not biased in all of the cases. Figure~\ref{fig:combine150s15}
shows an example when both band measurement uncertainty and frequency dependent polarization rotation are 
considered at 150~GHz band. When the 150~GHz band-center or band-width is mis-estimated by 15 GHz, the 
estimated scaling coefficients and CMB $B$-mode signal are without any bias. Similarly, the formalism does 
not bias the CMB $B$-mode signal with a mis-estimation of 15 GHz of the band-center or band-width at 250 or 
410~GHz band. We also see similar results when only one of the systematic effects is considered.

\subsection{Band Measurement Uncertainty}
\label{sec:maxlikeresultbandshape}

When estimating the CMB $B$-mode signal in the presence of band 
measurement uncertainty and instrumental noise, all the band averaged rotation angles are assumed to be 
perfectly known. 

\begin{figure}[!h]
   \begin{center}
     \scalebox{0.5}{\includegraphics{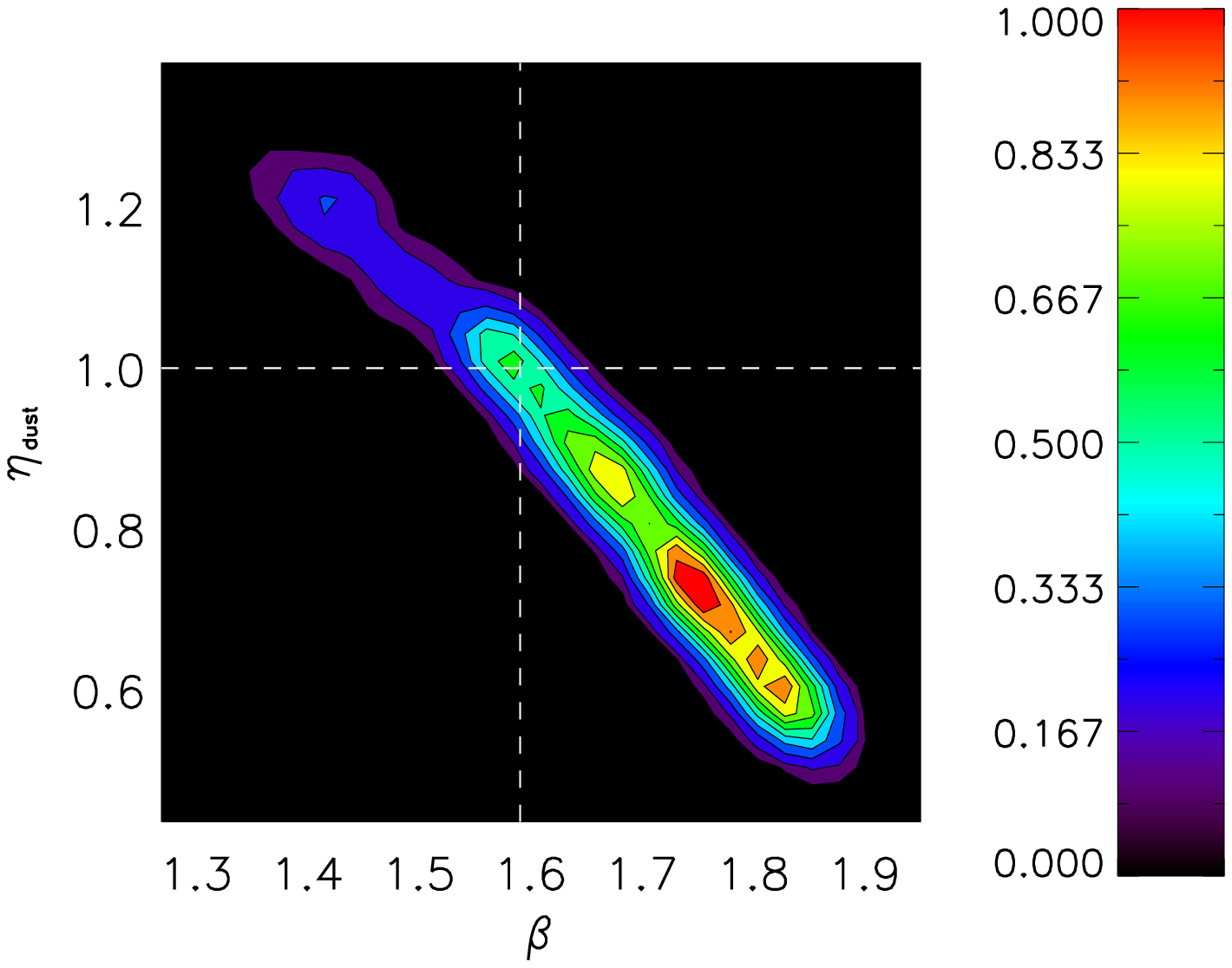}}
     \scalebox{0.35}{\includegraphics{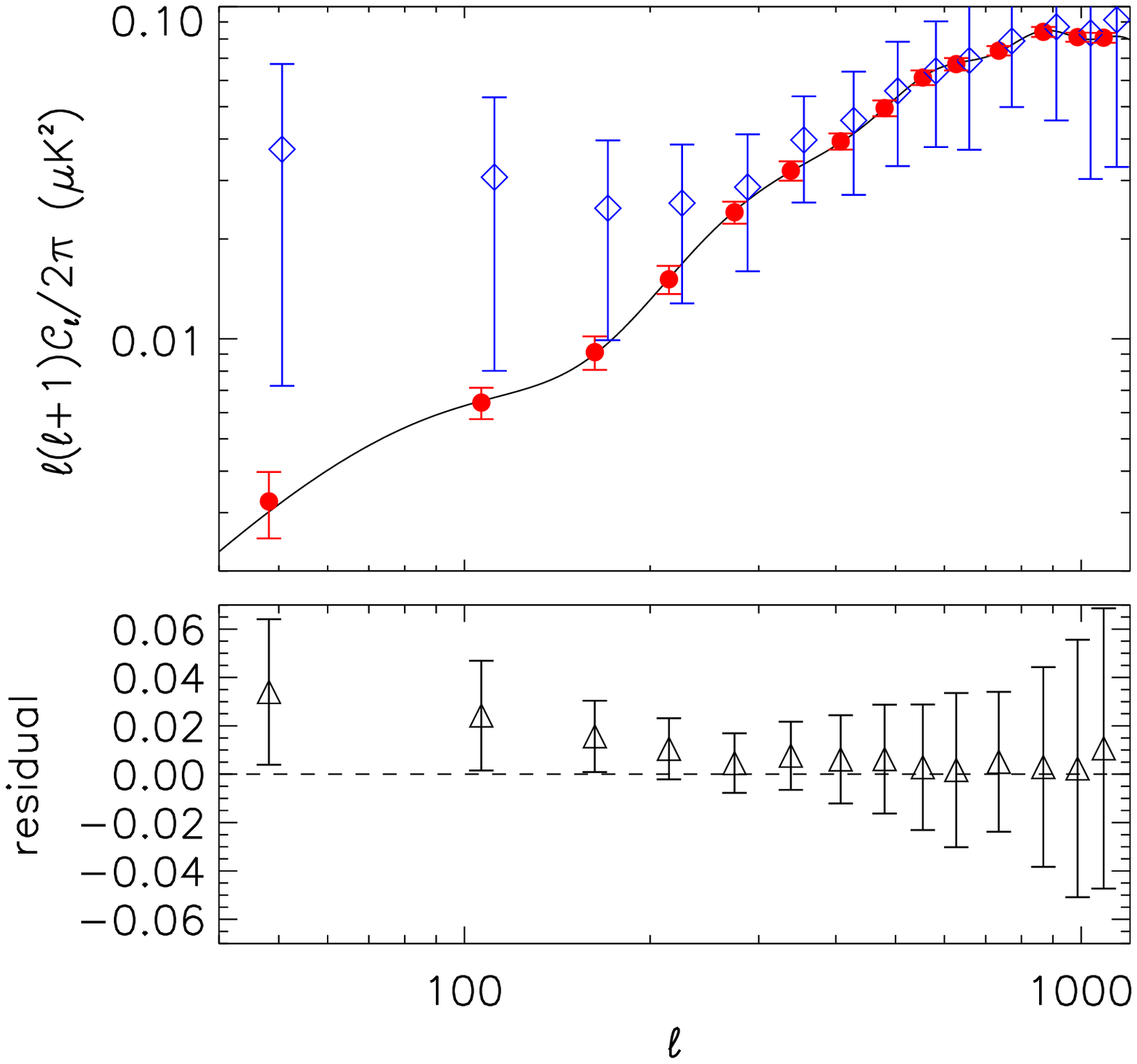}}
     \caption{Degeneracy between spectral index and dust scaling coefficient exists and it causes bias in 
       the estimated CMB $B$-mode power spectrum at low $\ell$. The scaling coefficients at 250 GHz and 
       410 GHz bands are assumed to be perfectly known. Only the two coefficients at 150 GHz are being 
       optimized. Left: 2-D likelihood plot between the dust spectral index $\beta$ and the dust scaling 
       coefficient $\eta_{\textrm{d}}$ at 150 GHz. The bright diagonal feature in the 2-D parameter space 
       shows that points along the strip give similar likelihood and thus are degenerate. The two white 
       dashed lines are the input values of the two parameters. The input values, although on the diagonal 
       feature, is not the point with the highest likelihood in this simulation. Right: Power spectrum (top) 
       and residual power spectrum (bottom) of the CMB $B$-mode. The estimated CMB signal~(blue diamond) is 
       biased compared to the input~(red dot) and the residual~(black triangle) is biased.}
     \label{fig:banddegebias}
   \end{center}
 \end{figure}

First we address the cases where there are uncertainties in only one of the bands. As discussed in 
Sec.~\ref{sec:addbandeffect} we observe a degeneracy between the dust spectral index and the dust scaling 
coefficient. When noise is present in the observation, changing the tilt of the dust spectrum or the 
scaling of the dust signal in one band give similar likelihood. The degeneracy causes bias in the recovered 
CMB power spectrum, particularly in the low-$\ell$ bins where the dust signal is high and the inflationary 
$B$-mode signal resides. Figure~\ref{fig:banddegebias} shows the 2-D likelihood plot between the dust 
spectral index and the dust scaling coefficient and the estimated CMB $B$-mode power spectrum when there are 
uncertainties at 150 GHz band only. The points along the bright diagonal feature in the likelihood plot have 
similar likelihood. When the parameters are mis-estimated, the estimated $B$-mode signal is biased by at 
low $\ell$.

\begin{figure}[!h]
 \begin{center}
 \scalebox{0.5}{\includegraphics{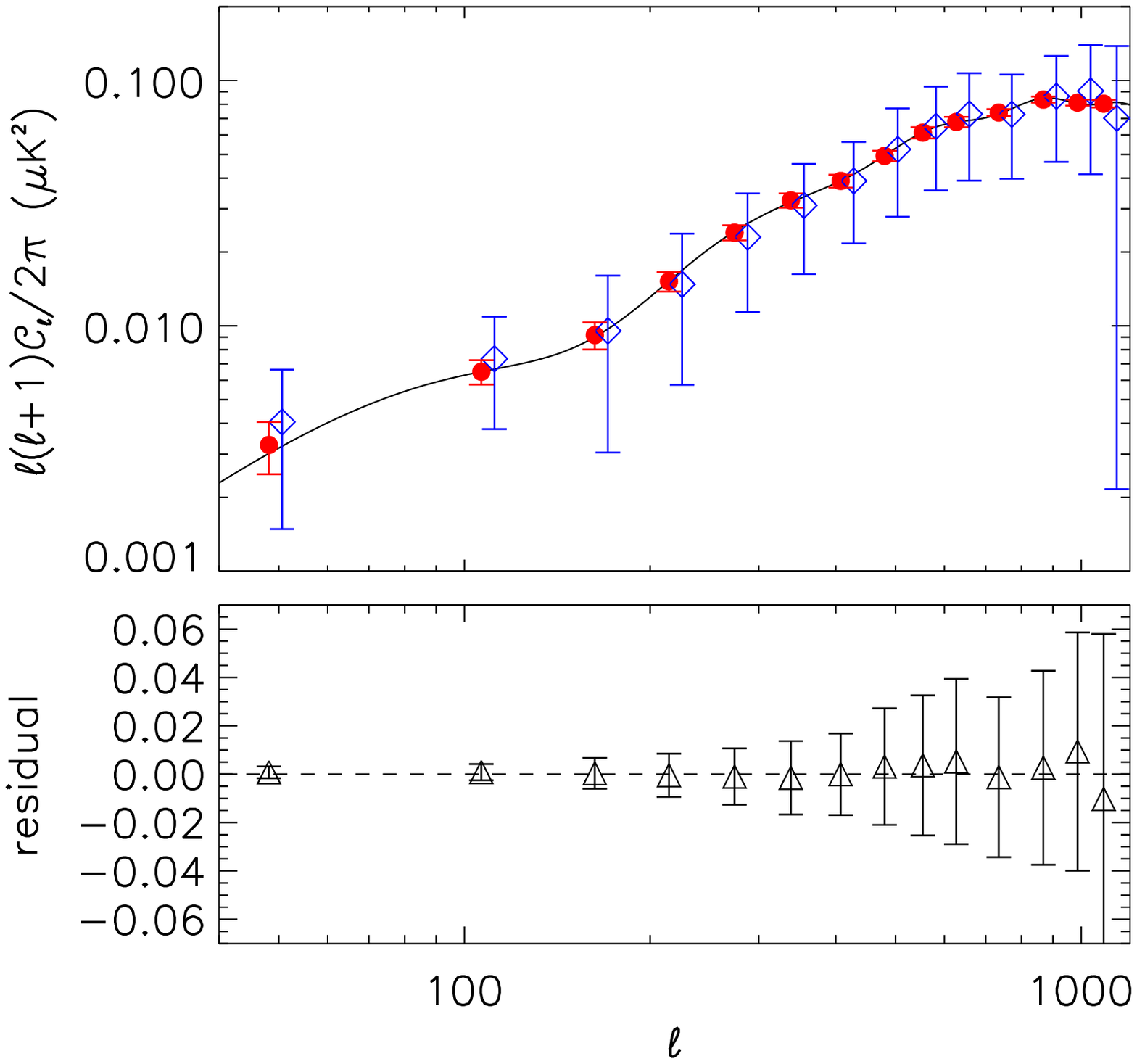}}
 \caption{Simulated results with 15\% priors on scaling coefficients at 150 GHz band. The coefficients at 
   250~GHz and 410~GHz bands are assumed to be perfectly known. The estimated CMB signal~(blue diamond) 
   agrees with the input~(red dot) and the residual~(black triangle) is consistent with zero.}
 \label{fig:150prior15}
 \end{center}
 \end{figure}

Setting priors on the scaling coefficients according to the band measurement uncertainty limits the 
parameter space in which the best fit values are searched for. In this study the priors are centered on the 
input value which is one. In the case where the uncertainty of only the 150 GHz band is considered, with 
15\% Gaussian priors on the scaling coefficients for both CMB and dust the estimated CMB $B$-mode signal is 
not biased, as shown in Fig.~\ref{fig:150prior15}. 

\begin{figure}[!h]
  \plottwo{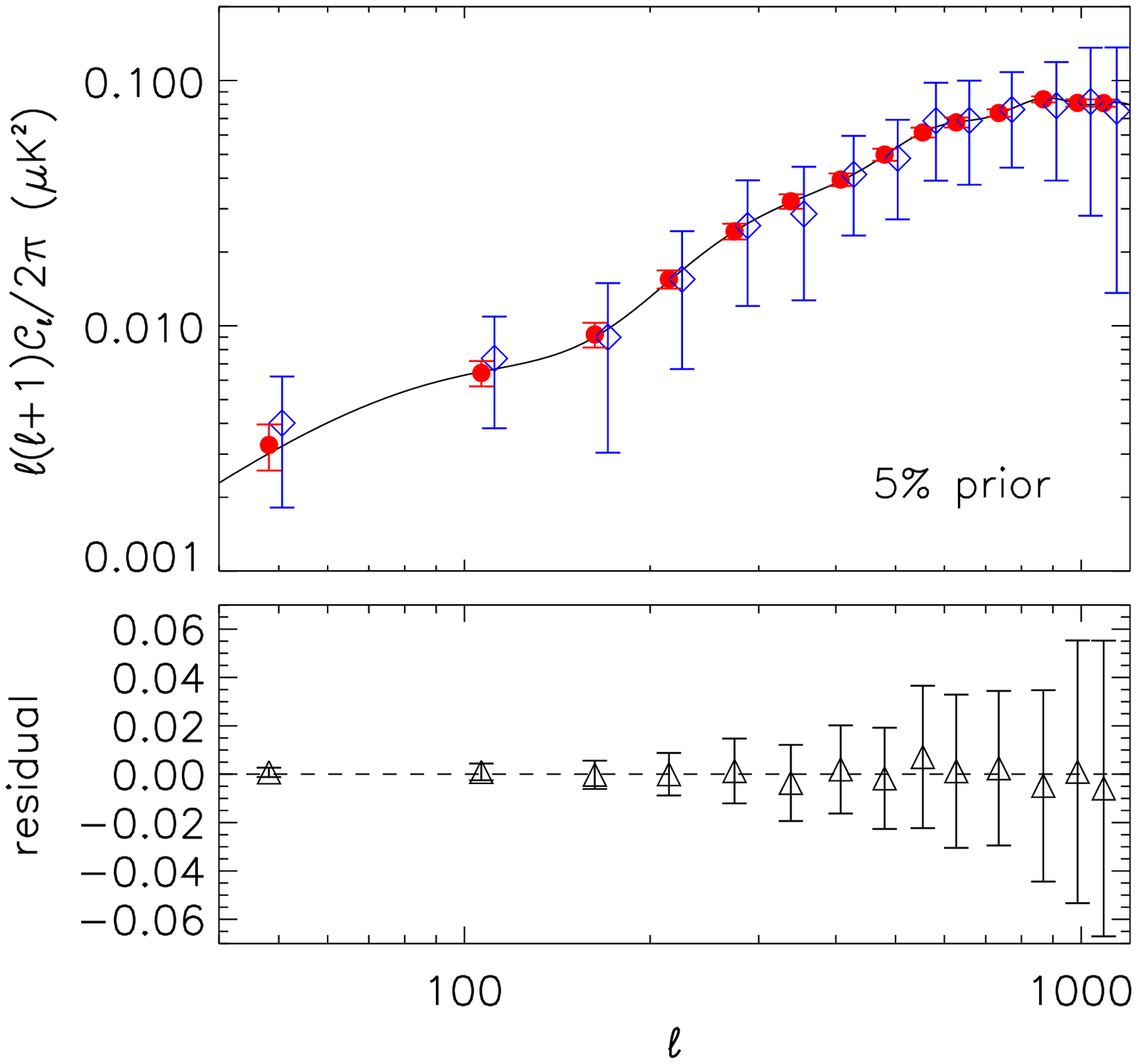}{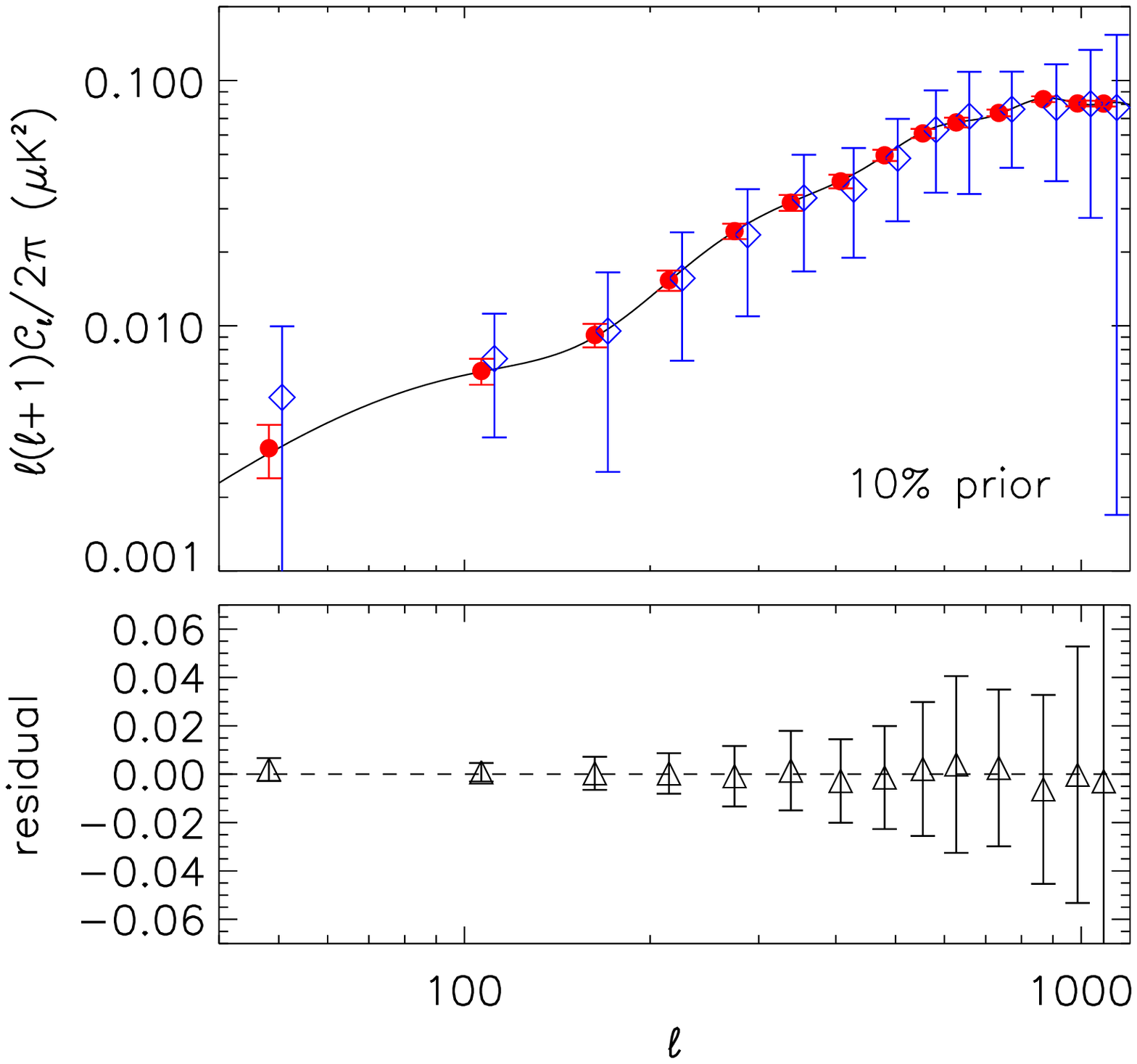}
    \caption{Simulation results where band measurement uncertainties of all three frequency bands are 
      included. 5\% (left) or 10\% (right) Gaussian priors are set on all scaling coefficients around the 
      input value. When the priors are 5\% the estimated CMB $B$-mode signal~(blue diamond) agrees with the 
      input CMB signal~(red dot). When the priors are relaxed to 10\%, the estimated $B$-mode signal at the 
      lowest-$\ell$ bin is 2.5 times the cosmic variance away from the input value. With the effect of 
      instrumental noise, the residual power spectrum~(black triangle) is consistent with zero.}
    \label{fig:bandallbandprior}
\end{figure}

When uncertainties of all three bands are included the degeneracy between the scaling coefficients and
the sky signal level requires priors on all scaling coefficients. Figure~\ref{fig:bandallbandprior} shows 
two cases where 5\% or 10\% Gaussian priors are set on all scaling coefficients. When all scaling 
coefficients have 5\% Gaussian priors the $B$-mode signal is estimated accurately. When the priors 
are relaxed to 10\%, the estimated signal in the lowest $\ell$ bin is 2.5 times the cosmic 
variance away from the input value. The bias comes from the degeneracy between 
the dust scaling coefficients and the dust spectral index and therefore the bias
is strongest at the $\ell$ range in which dust is most intense. If we factor in the 
effect of the instrumental noise the residual power spectrum is consistent with zero in all $\ell$ bins. 

\begin{figure}[!h]
  \begin{center}
    \scalebox{0.7}{\includegraphics{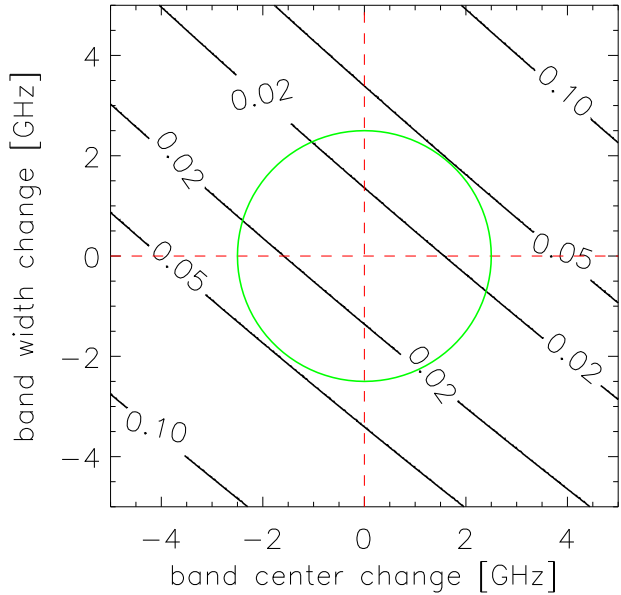}}
    \caption{Contours showing percent scaling coefficient deviations from the nominal value of unity 
      translated to band-center and band-width shifts from their nominal value for 
     the 250~GHz band for Galactic dust. The green circle shows 
      the constraint imposed on the scaling coefficient 
      with 2.5 GHz uncertainty for both band-center and band-width. The circle 
      lies entirely within the 5\% contour of scaling coefficient deviation implying that 2.5~GHz band
      measurement uncertainty in this band is equivalent to a 5\% prior on $\eta_{\textrm{d}}$.}
    \label{fig:eta_contour}
  \end{center}
\end{figure}

Specific priors on the scaling coefficients is analogous to known measurement uncertainties. 
Table~\ref{tab:etatobandmismatch} gives the mapping between prior interval on the 
scaling coefficients and uncertainties in band-center and band-width for the three top-hat bands used in the 
simulations. The values for band-center uncertainty are calculated assuming the band-width has no uncertainty 
and vice versa. 
\begin{table}[!h]
\begin{center}
\begin{tabular}{|c|c|c|c|c|}
\hline
 & Band mismatch& 150 GHz & 250 GHz & 410 GHz \\\hline
\multirow{2}{*}{5\% $\eta_{\textrm{d}}$ mismatch} & center (GHz) & 4.5 & 7.5 & 12.0 \\
 & width (GHz) & 2.0 & 3.5 & 4.0 \\\hline
\multirow{2}{*}{10\% $\eta_{\textrm{d}}$ mismatch} & center (GHz) & 9.5 & 15.0 & 25.0 \\
 & width (GHz) & 4.0 & 7.0 & 8.5 \\\hline
\end{tabular}
\caption{The band-center and band-width mismatches corresponding to a 5\% or 10\% $\eta_{\textrm{d}}$ mismatch 
  for the three top-hat bands at 150, 250 and 410~GHz. When calculating the values for band-center, the 
  band-widths are assumed to be perfectly known and vice versa.}
\label{tab:etatobandmismatch}
\end{center}
\end{table}
In practice, however, both band-center and band-width have uncertainty simultaneously. As an example, in 
Fig.~\ref{fig:eta_contour} we show the contour of the scaling coefficient deviation in the 2-D parameter 
space of band-center and band-width for Galactic dust at 250 GHz band. Similar 
Figures are obtained for the CMB and other frequency bands as well. Using such Figures, 
which are easily calculable given any band parameters, one can map band measurement uncertainty
to scaling coefficient priors.  

\subsection{Frequency Dependent Polarization Rotation}
\begin{figure}[h]
  \begin{center}
    \scalebox{0.5}{\includegraphics{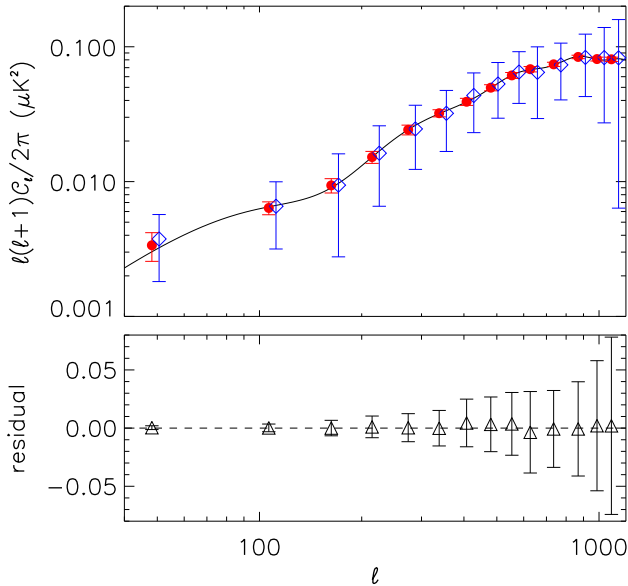}}
    \caption{Simulation results where band averaged polarization rotation angles for only 150 GHz band are 
      optimized. No prior constraints are set on the rotation angles at 150 GHz. All angles at 250 GHz and 
      410 GHz bands are assumed to be perfectly known. The estimated CMB $B$-mode signal~(blue diamond) 
      agrees with the input signal~(red dot). 
    }
 \label{fig:ahwp150noisefitrot}
 \end{center}
 \end{figure}

When assessing only the frequency dependent polarization rotation in the presence of instrumental noise, all 
the scaling coefficients are assumed to be perfectly known. We first consider the uncertainty of the 
band averaged rotation angles in only one band while assuming the rotation angles in the other two bands are 
perfectly known. Since there is no degeneracy between the incoming polarization angle and the AHWP
rotation angles in this case, we can estimate the CMB $B$-mode signal accurately without priors on the 
rotation angles. Figure~\ref{fig:ahwp150noisefitrot} shows the result where the uncertainty of the 
polarization angle calibration in \textit{only} 150~GHz band is considered. We get similar results for 
250~GHz and 410~GHz band.

\begin{figure}[h]
  \plottwo{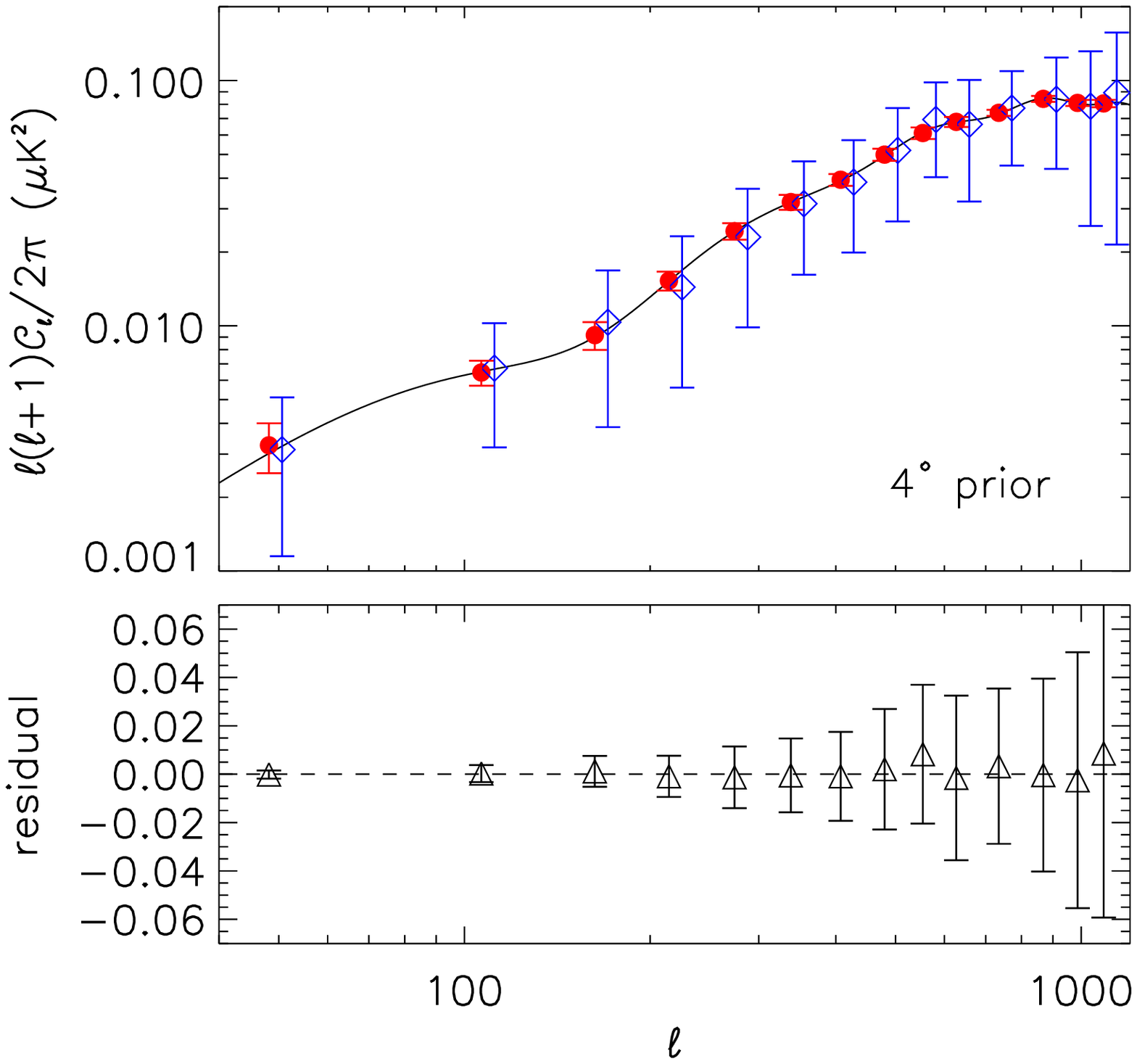}{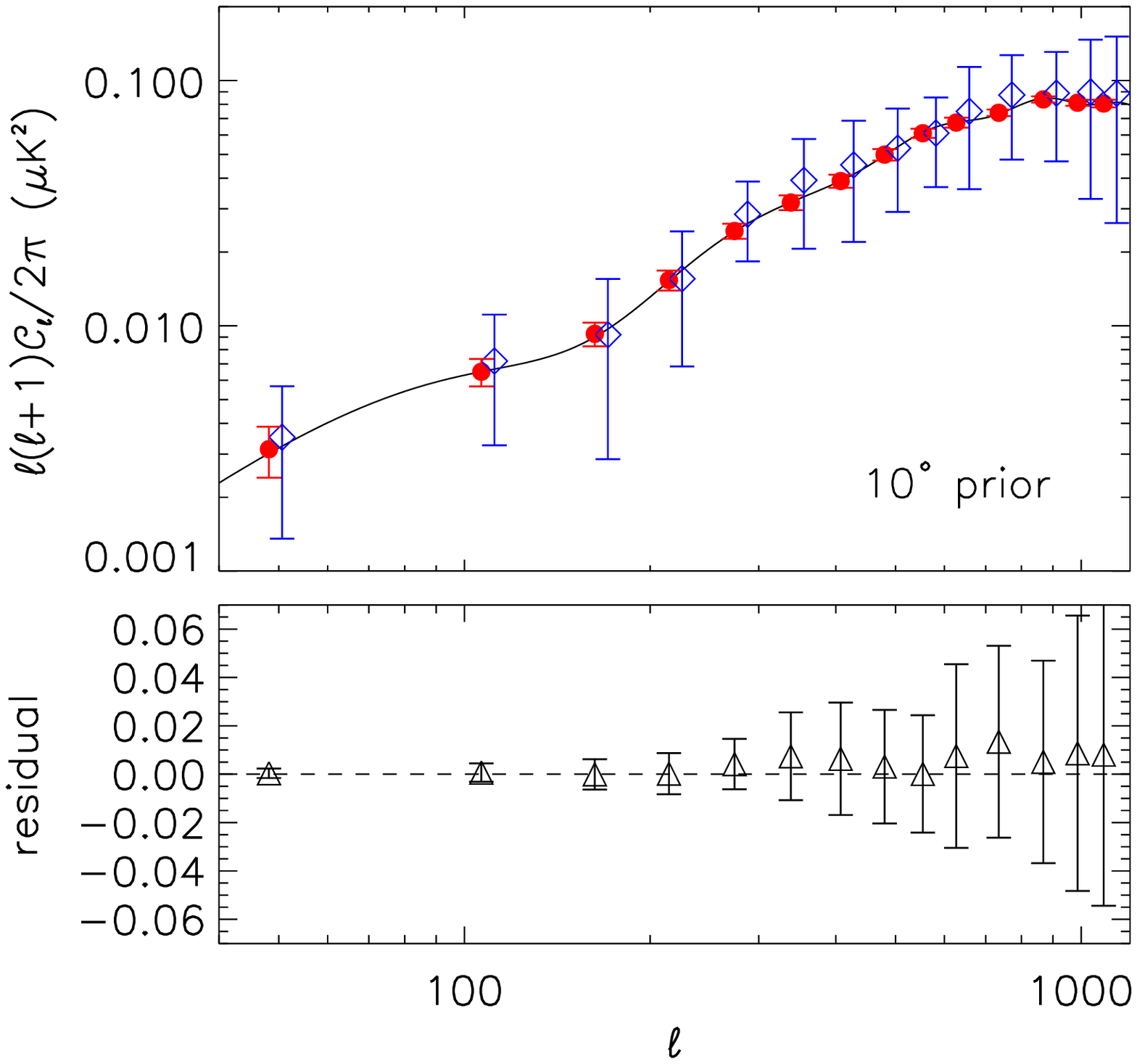}
    \caption{Simulation results where the uncertainties of polarization angle calibration for all three 
      bands are included. Left: 4$^{\circ}$~Gaussian priors on all rotation angles. Right: 
      10$^{\circ}$~Gaussian priors on all rotation angles. When the priors are 4$^{\circ}$\ the estimated CMB 
      $B$-mode signal~(blue diamond) agrees with the input signal~(red dot). When the priors are 
      10$^{\circ}$, the estimated CMB $B$-mode is more than twice the cosmic variance away from the input CMB 
      in a few bins at $\ell > 300$ due to the mixing of $E$-mode signal into $B$-mode signal. When the 
      instrumental noise is considered, the residual power spectrum~(black triangle) is consistent with 
      zero.}
 \label{fig:ahwpallbandprior}
 \end{figure}

When uncertainties of polarization angle calibration in all three bands are included, the degeneracy 
between the frequency dependent polarization rotation and the polarization angle of the incoming signal 
requires priors on the band-averaged rotation angles. Here we center the priors on the input value. 
Figure~\ref{fig:ahwpallbandprior} shows that with 4$^{\circ}$ Gaussian priors on all band-averaged rotation 
angles, the final estimated CMB $B$-mode signal is not biased. When the priors are relaxed to 10$^{\circ}$ 
the estimated CMB $B$-mode signal is more than twice the cosmic variance away from the input in a few bins 
at $\ell > 300$. Less stringent priors on the band averaged rotation angles result in a bigger mis-estimate 
of the polarization angle of the incoming signal. This mis-estimate induces leakage from CMB $E$-mode signal 
to $B$-mode signal which has a bigger effect at high $\ell$. When other noise sources are included 
the residual power spectrum is consistent with zero in both cases.

\subsection{Combining Band Measurement Uncertainty and Frequency Dependent Polarization
  Rotation}
  
\begin{figure}[!h]
  \begin{center}
    \scalebox{0.5}{\includegraphics{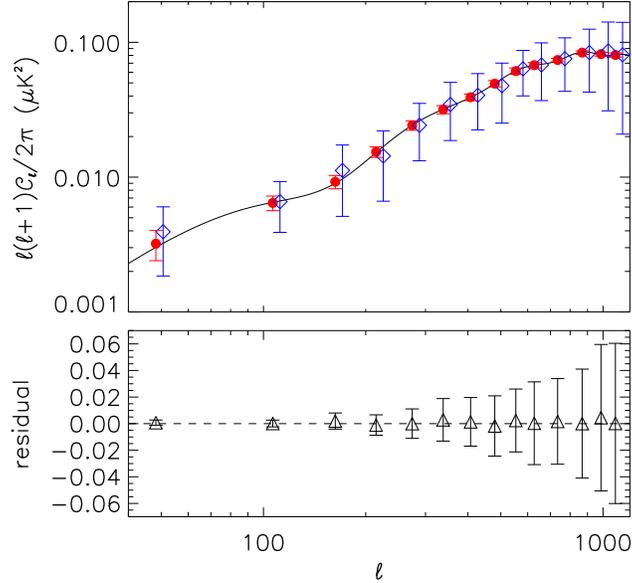}}
    \caption{Simulation results where both the band measurement uncertainty and frequency dependent 
      polarization rotation are included for the 150 GHz band in the presence of instrumental
      noise. There are 15\% Gaussian priors on the scaling coefficients and no prior on the band-averaged 
      rotation angles. The parameters for the 250 GHz and 410 GHz bands are assumed to be perfectly known. 
      There is no bias in the estimated CMB $B$-mode signal~(blue diamond) compared to the input signal~(red 
      dot) and the residual~(black triangle) is consistent with zero.
    }
 \label{fig:combine150noiseprior}
 \end{center}
 \end{figure}

\begin{figure}[!h]
  \begin{center}
    \scalebox{0.5}{\includegraphics{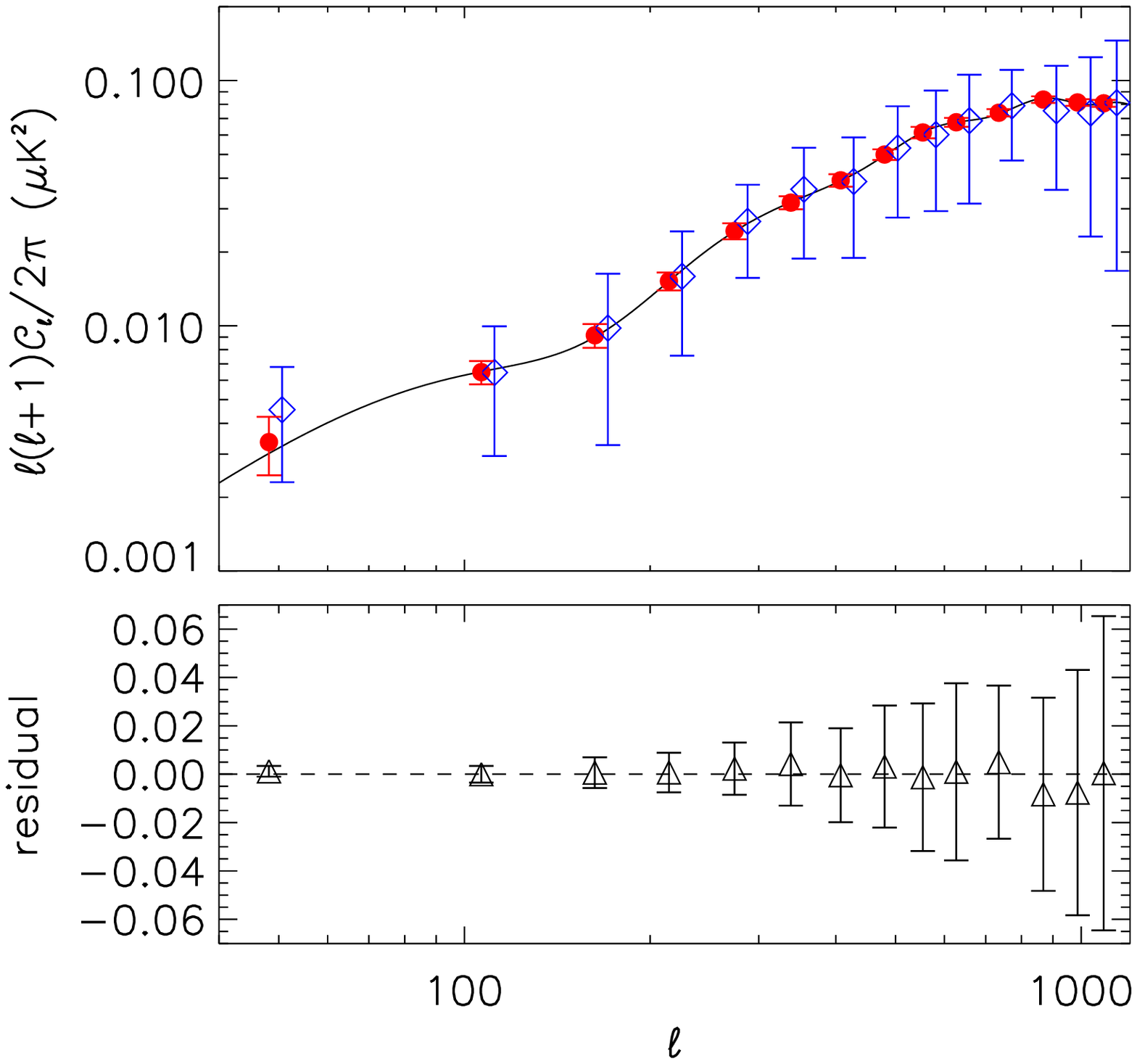}}
    \caption{Results of simulation with 5\% Gaussian priors on scaling coefficients and 4$^{\circ}$ Gaussian 
      priors on band averaged rotation angles in the presence of instrumental noise. The estimated CMB 
      signal~(blue diamond) is not biased compared to the input~(red dot) and the residual~(black triangle) 
      is consistent with zero.}
 \label{fig:noiseband5periva2deg}
 \end{center}
 \end{figure}

When both systematic effects are included in the presence of instrumental noise, the formalism estimates the 
scaling coefficients and the band averaged polarization rotation angles simultaneously. We first assess the 
bias in the estimated CMB $B$-mode signal including the systematic effects at only one of the bands while 
assuming the other two bands are perfectly known. In Fig.~\ref{fig:combine150noiseprior} we show the results 
where there are systematic effects at 150 GHz only. The estimated CMB $B$-mode signal is not biased with 
15\% Gaussian priors on the scaling coefficients for CMB and dust and no priors on band averaged rotation 
angles. We get similar results when there are systematic effects at 250 or 410~GHz band only.
When the systematic effects are considered for all three bands, priors are needed for all parameters due to 
the degeneracies. Figure~\ref{fig:noiseband5periva2deg} shows that with 5\% Gaussian prior on scaling 
coefficients and 4$^{\circ}$ Gaussian priors on band averaged rotation angles we can estimate CMB $B$-mode 
signal without any bias.

\subsection{Foreground Estimation with Measured EBEX Bands}

\begin{figure}[!h]
  \begin{center}
    \scalebox{0.5}{\includegraphics{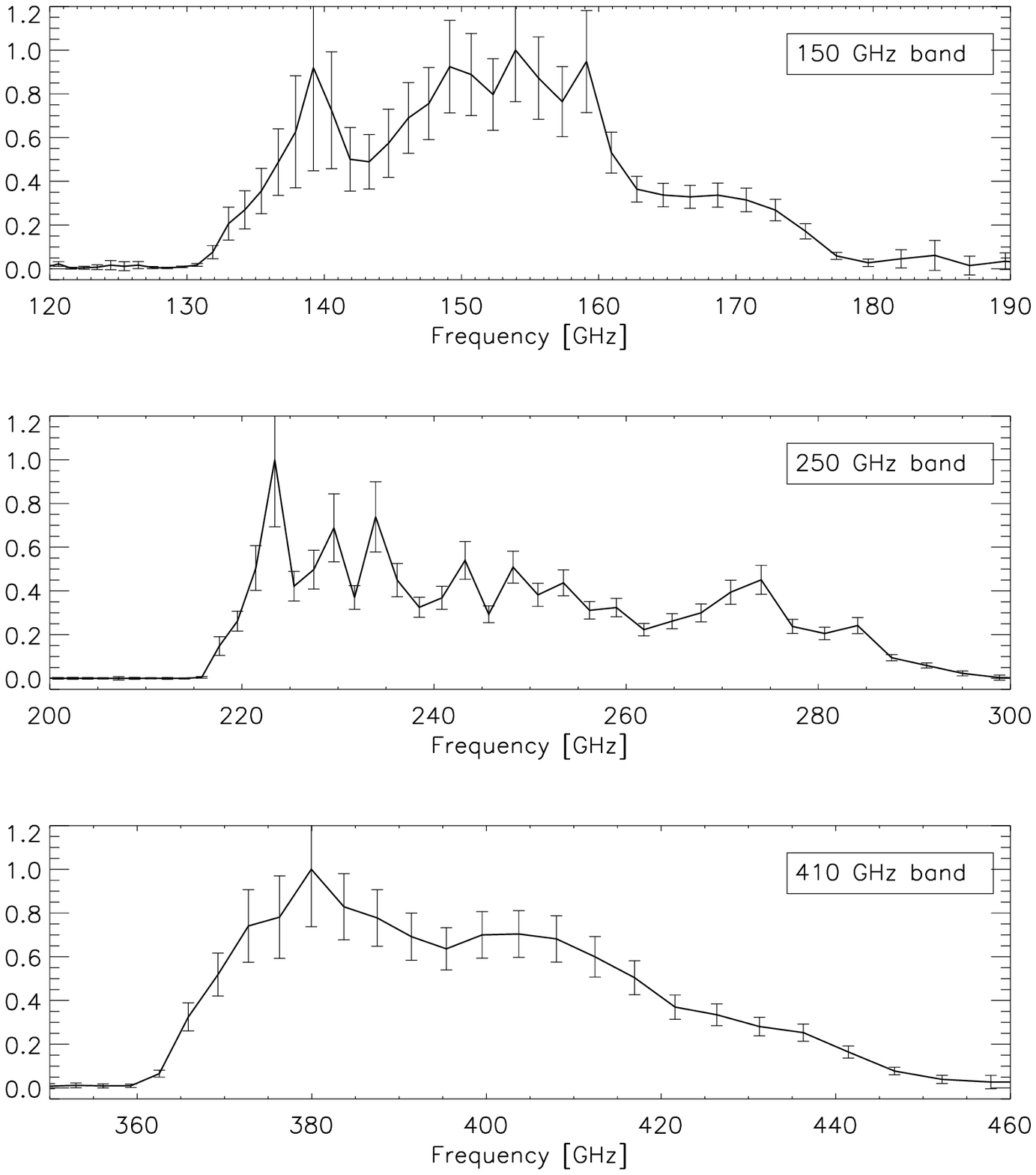}}
    \caption{Measured EBEX 150 GHz, 250 GHz and 410 GHz bands. The measurements and 
      data analysis to derive these bands are presented in \citet{zilicthesis}.}
    \label{fig:ebexldbbands}
  \end{center}
\end{figure}

As a practical example, we estimate foreground using the measured EBEX bands and their uncertainties, which 
are shown in Fig.~\ref{fig:ebexldbbands} \citep{zilicthesis}. 
We carry out simulations of the band measurement. 
Each realization is a simulated measurement of the band given the measured frequency bin errors.
For each realization we compute the scaling coefficients compared to the measured 
band, which are taken to be the nominal values of the measurement, 
and band averaged rotation angles for CMB and Galactic dust. The standard deviation from the simulated 
results, listed in Table~\ref{tab:ldbbandrotpriors}, are considered as the priors 
on the corresponding parameters.
\begin{table}[!h]
\begin{center}
\begin{tabular}{|c|c|c|c|}
\hline
Priors & 150 GHz band & 250 GHz band & 410 GHz band \\ \hline
$\eta_{CMB}$ & 5\% & 4\% & 5\% \\\hline
$\eta_{dust}$ & 5\% & 3\% & 4\% \\\hline 
$\theta_{CMB}$ & 0.2$^{\circ}$ & 0.02$^{\circ}$ & 0.2$^{\circ}$ \\\hline
$\theta_{dust}$ & 0.2$^{\circ}$ & 0.02$^{\circ}$ & 0.2$^{\circ}$ \\\hline
\end{tabular}
\caption{Uncertainties of the in-band scaling coefficients $\eta$ and band averaged rotation angles $\theta$ 
  for the three EBEX bands. These values are the standard deviation of the parameters calculated from 500 
  Monte-Carlo simulations.}
\label{tab:ldbbandrotpriors}
\end{center}
\end{table}

\begin{figure}[!h]
  \begin{center}
    \scalebox{0.5}{\includegraphics{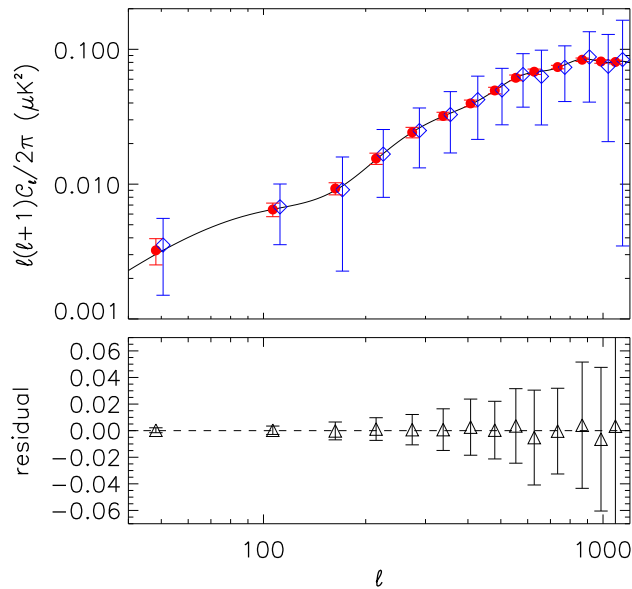}}
    \caption{
      Simulation results where both systematic effects in all three bands are included. All parameters have 
      priors based on Monte-Carlo simulations using the three EBEX bands, which are listed in 
      Table~\ref{tab:ldbbandrotpriors}. The estimated CMB $B$-mode signal (blue diamond) agree with the 
      input signal (red dot) without any bias and the residual (black diamond) is consistent with zero.}
 \label{fig:maxlikefitbandrotldb}
 \end{center}
 \end{figure}

Figure~\ref{fig:maxlikefitbandrotldb} shows the power spectra of the simulation with the priors
listed in Table~\ref{tab:ldbbandrotpriors}. We find that with $r = 0.05$ an expreimtn like EBEX can 
estimate the B-mode spectrum without bias at a level comparible to either cosmic variance or instrumental
noise.

\section{Summary}
\label{sec:maxlikediscussion}

Recent observations suggest that Galactic 
dust is a significant contaminating source in all regions of the sky for CMB polarimeters 
targeting the inflationary $B$-mode signal. The Galactic dust signal needs to be removed and robust 
foreground estimation is essential. We presented a general maximum likelihood foreground estimation 
formalism in the presence of two systematic effects: band measurement uncertainty and frequency dependent 
polarization rotation. The formalism fits the systematic effects simultaneously with the dust spectral index 
which allows for imperfect knowledge of the instrumental parameters. 

We found several degeneracies: 1) between the dust spectral index and the dust scaling coefficients, 
2) between the scaling coefficients and the signal levels, and 3) between the frequency dependent 
polarization rotation and the signal polarization angles. Due to the degeneracies, priors on the parameters
are needed in order estimate the CMB $B$-mode signal accurately. We quantified the degeneracies
and showed, as an example, that a sub-orbital experiment like EBEX should not be limited in estimating and 
subtracting Galactic dust if it has 10\% polarization fraction, if the tensor to scalar ratio $r = 0.05$, 
and with band measurement and polarization angle calibration uncertainty of 5\% and 4$^{\circ}$, 
respectively. Such an experiment may not be limited with even higher polarization fractions or 
lower $r$ values. A detailed calculation is necessary for other concrete cases. 

Band measurement uncertainty is a common systematic effect for all CMB instruments. 
Frequency dependent polarization rotation by elements in the focal plane is not unique to the use of an 
achromatic half-wave plate either. Sinuous antenna detectors also have frequency dependent 
polarization rotation \citep{obrient2008b}. Therefore, 
the foreground estimation formalism developed in this paper has general applicability.

\acknowledgments
This work received major support through an NSF grant NSF ANT-094513. 
 We are thankful for the computing resources provided by the Minnesota Supercomputing
 Institute. EBEX is supported through NASA grants NNX08AG40G and NNX07AP36H. C.Bao acknowledges the support
 from the Doctoral Dissertation Fellowship from the University of Minnesota. C.Baccigalupi acknowledges the
 support from the INDARK INFN Grant.

\bibliography{cybao}
\bibliographystyle{apj}
\end{document}